\documentclass[12pt]{article}
\usepackage{epsfig}

\def\br(#1,#2){\left\langle#1#2\right\rangle}
\def\sq(#1,#2){\left[#1#2\right]}
\def\s(#1,#2){s_{#1 #2}}
\def\t(#1,#2,#3){s_{#1 #2 #3}}

\begin{document}
\begin{titlepage}

\hspace*{\fill}\parbox[t]{5cm}
{hep-ph/0204093 \\
ANL-HEP-PR-02-027 \\
FERMILAB-Pub-02/062-T \\
ILL-(TH)-02-3 \\ \\
\today} \vskip2cm
\begin{center}
{\Large \bf Higgs-Boson Production in Association with a\\
\medskip Single Bottom Quark} \\
\medskip
\bigskip\bigskip\bigskip\bigskip
{\large  {\bf J.~Campbell}$^1$,
         {\bf R.~K.~Ellis}$^2$,
         {\bf F.~Maltoni}$^3$,
     and {\bf S.~Willenbrock}$^3$} \\
\bigskip\bigskip\medskip
$^{1}$High Energy Physics Division,
Argonne National Laboratory \\
Argonne, IL\ \ 60439 \\ \bigskip
$^{2}$Theoretical Physics Department, Fermi National Accelerator Laboratory \\
P.~O.~Box 500, Batavia, IL\ \ 60510 \\ \bigskip
$^{3}$Department of Physics, University of Illinois at Urbana-Champaign \\
1110 West Green Street, Urbana, IL\ \ 61801 \\ \bigskip
\end{center}

\bigskip\bigskip\bigskip

\begin{abstract}
Higgs bosons from an extended Higgs sector, such as a two-Higgs-doublet model,
can have greatly enhanced coupling to the bottom quark.  Producing such a Higgs
boson in association with a single high-$p_T$ bottom quark via $gb\to hb$
allows for the suppression of backgrounds.  Previous studies have instead used
$gg,q\bar q\to b\bar bh$ as the production mechanism, which is valid only if
both $b$ quarks are at high $p_T$.  We calculate $gb\to hb$ at next-to-leading
order in QCD, and find that it is an order of magnitude larger than $gg,q\bar
q\to b\bar bh$ at the Fermilab Tevatron and the CERN Large Hadron Collider.
This production mechanism improves the prospects for the discovery of a Higgs
boson with enhanced coupling to the $b$ quark.
\end{abstract}

\end{titlepage}

\section{Introduction}%
\label{sec:intro}

The Higgs boson couples to fermions with strength $m_f/v$, where $v=(\sqrt 2
G_F)^{-1/2} \approx 246$ GeV is the vacuum expectation value of the Higgs
field. Its Yukawa coupling to bottom quarks ($m_b \approx 5$ GeV) is thus very
weak, leading to very small cross sections for associated production of the
Higgs boson and bottom quarks at the Fermilab Tevatron \cite{Stange:ya} and
the CERN Large Hadron Collider (LHC) \cite{Dicus:1988cx}.  However, this Yukawa
coupling could be considerably enhanced in extensions of the standard model
with more than one Higgs doublet, thereby increasing this production cross
section \cite{Dicus:1988cx}.  For example, in a two-Higgs-doublet model, the
Yukawa coupling of some or all of the Higgs bosons ($h^0,H^0,A^0,H^\pm$) to the
bottom quark could be enhanced for large values of $\tan\beta = v_2/v_1$,
where $v_1$ is the vacuum expectation value of the Higgs doublet that couples
to the bottom quark.

\begin{figure}[ht]
\begin{center}
\vspace*{0cm} \hspace*{0cm} \epsfxsize=4cm \epsfbox{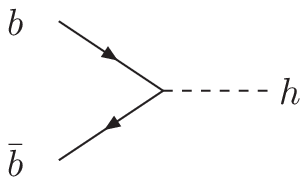} \vspace*{-.8cm}
\end{center}
\caption{Production of the Higgs boson via $b\bar b\to h$.  There are
typically no high-$p_T$ bottom quarks in the final state.} \label{fig:bbh}
\end{figure}

The dominant subprocess for the production of a Higgs boson via its coupling to
bottom quarks is $b\bar b\to h$ (Fig.~\ref{fig:bbh}),\footnote{We use $h$ to
denote a generic Higgs boson.  In a two-Higgs-doublet model, $h$ may denote
any of the neutral Higgs bosons ($h^0,H^0,A^0$).} where the $b$ quarks reside
in the proton sea \cite{Dicus:1988cx,Dicus:1998hs}.  The $b$-quark sea is
generated from gluons splitting into nearly collinear $b\bar b$ pairs. When
one member of the pair initiates a hard-scattering subprocess, its partner
tends to remain at low $p_T$ and to become part of the beam remnant. Hence the
final state typically has no high-$p_T$ bottom quarks.  This subprocess may be
useful to discover a Higgs boson for large $\tan\beta$ in the decay mode $h\to
\tau^+\tau^-$ at the Tevatron and the LHC \cite{Kunszt:1991qe,unknown:1999fr},
and $h\to \mu^+\mu^-$ at the LHC
\cite{unknown:1999fr,Kao:1995gx,Barger:1997pp}.  The decay mode $h\to b\bar b$
is not distinguishable from the overwhelming background $gg,q\bar q\to b\bar
b$.

\begin{figure}[ht]
\begin{center}
\vspace*{0cm} \hspace*{0cm} \epsfxsize=10cm \epsfbox{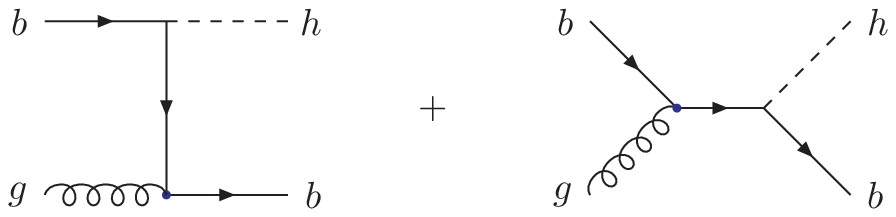} \vspace*{-.8cm}
\end{center}
\caption{Associated production of the Higgs boson and a single high-$p_T$
bottom quark.} \label{fig:gbhb}
\end{figure}

If one instead demands that at least one $b$ quark be observed at high $p_T$,
then the leading-order subprocess for associated production of the Higgs boson
and bottom quarks is $gb\to hb$ (Fig.~\ref{fig:gbhb})
\cite{Choudhury:1998kr,Huang:1998vu}.\footnote{This includes the
charge-conjugate subprocess $g\bar b \to h\bar b$.  All charge-conjugate
subprocesses are understood throughout this paper.} The presence of a
high-$p_T$ bottom quark in the final state has distinct phenomenological
advantages since it can be tagged with reasonably high efficiency. In the case
of $h \to \tau^+\tau^-,\mu^+\mu^-$ the $b$ quark can be used to reduce
backgrounds and to identify the Higgs-boson production mechanism
\cite{unknown:1999fr,Drees:1997sh,Carena:1998gk}.  The trade-off is that the
cross section for $gb\to hb$, with the $b$ quark at high $p_T$, is less than
that of $b\bar b\to h$.

\begin{figure}[ht]
\begin{center}
\vspace*{0cm} \hspace*{0cm} \epsfxsize=10cm \epsfbox{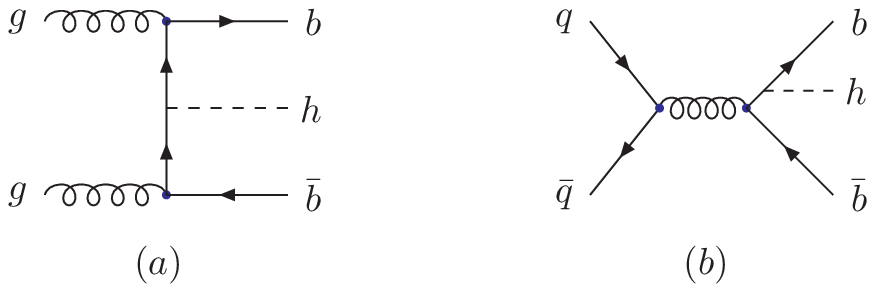} \vspace*{-.8cm}
\end{center}
\caption{Representative diagrams for associated production of the Higgs boson
and two high-$p_T$ bottom quarks: (a) $gg\to b\bar bh$ (8~diagrams); (b) $q\bar
q\to b\bar bh$ (2~diagrams).} \label{fig:ggbbh}
\end{figure}

If the Higgs boson decays via $h\to b\bar b$, the presence of an additional
high-$p_T$ bottom quark in the final state is essential in order to separate
the signal from backgrounds \cite{Dai:1994vu,Dai:1996rn}. Recent analyses are
based on the subprocess $gg,q\bar q\to b\bar bh$ (Fig.~\ref{fig:ggbbh}), and
demand a final state with four jets, with either at least three $b$ tags, or
with four $b$ tags \cite{unknown:1999fr,Carena:1998gk,Richter-Was:1997gi,
Diaz-Cruz:1998qc,Balazs:1998nt,Carena:2000yx}. However, the cross section for
this subprocess is less than that of $gb \to hb$.  We therefore suggest that
it may be advantageous to search for $h \to b\bar b$ by demanding just three
jets in the final state, all of which are $b$ tagged
\cite{Dai:1994vu,Dai:1996rn}.  The three-jet final state will have bigger
backgrounds than the four-jet final state, but the significance of the signal
($S/\sqrt B$) is likely to increase.

It is only valid to use $gg,q\bar q\to b\bar bh$ as the production subprocess
when both $b$ quarks are at high $p_T$.  If only one of the $b$ quarks is at
high $p_T$ \cite{unknown:1999fr,Dai:1994vu,Dai:1996rn}, the integration over
the momentum of the other $b$ quark yields a factor $\ln (m_h/m_b)$ which
invalidates perturbation theory. Our calculation of $gb\to hb$ sums these
logarithms to all orders, and results in a well-behaved perturbation series.

In this paper we calculate the cross section for the associated production of
the Higgs boson and a single bottom quark ($gb\to hb$) at next-to-leading
order. We provide results for both the Tevatron and the LHC. The cross section
for the subprocess $b\bar b\to h$ is already known at next-to-leading order
\cite{Dicus:1998hs,Balazs:1998sb}. The cross section for the subprocess
$gg,q\bar q\to b\bar bh$, which has two high-$p_T$ bottom quarks, is known
only at leading order, but the analogous subprocess $gg,q\bar q\to t\bar th$
has been calculated at next-to-leading order
\cite{Beenakker:2001rj,Reina:2001sf}, so the next-to-leading-order result for
$gg,q\bar q \to b\bar bh$ could be made available. Thus our calculation
completes the set of next-to-leading-order cross sections for the subprocesses
$b\bar b\to h$, $gb \to hb$, and $gg,q\bar q\to b\bar bh$.

In Section~\ref{sec:leading} we discuss the leading-order cross section for
$gb\to hb$.  In Section~\ref{sec:invlog} we discuss the correction of order
$1/\ln(m_h/m_b)$, due to initial gluons splitting into $b\bar b$ pairs.  In
Section~\ref{sec:alphas} we discuss the correction of order $\alpha_S$; the
virtual and real corrections are discussed separately.  We present our
numerical results in Section~\ref{sec:results}.  Conclusions are drawn in
Section~\ref{sec:conclusions}. Several appendices follow, in which the analytic
results and some of the technical details are presented.

\section{Leading order}%
\label{sec:leading}

The leading-order subprocess for Higgs-boson production in association with a
single high-$p_T$ bottom quark is shown in Fig.~\ref{fig:gbhb}. Since the scale
of the hard scattering is large compared with the $b$-quark mass, the $b$
quark is regarded as part of the proton sea
\cite{Olness:1987ep,Barnett:1987jw,Aivazis:1993pi,Collins:1998rz}.  However,
unlike the light-quark sea, the $b$-quark sea is perturbatively calculable.
This changes the way that one counts powers
\cite{Dicus:1998hs,Stelzer:1997ns}.  If the scale of the hard scattering is
$\mu$, the $b$ distribution function $b(x,\mu)$ is intrinsically of order
$\alpha_S(\mu)\ln(\mu/m_b)$, in contrast with the light partons, which are of
order unity.  This captures the behavior of the $b$ distribution function at
low and high values of $\mu$, and interpolates between them.  As $\mu$
approaches $m_b$ from above, $\ln(\mu/m_b)$ vanishes; this reflects the
initial condition on the $b$ distribution function, $b(x,m_b)=0$.  As $\mu$
becomes asymptotically large, $\alpha_S(\mu)\ln(\mu/m_b)$ approaches order
unity,\footnote{This can be seen by recalling $\alpha_S(\mu) \approx
2\pi/(\beta_0\ln(\mu/\Lambda_{QCD}))$.} and the $b$ distribution function
becomes of the same order as the light partons.

With this counting, the leading-order subprocess $gb\to hb$ is of order
$\alpha_S^2\ln(m_h/m_b)$ (times the Yukawa coupling), where we have chosen the
Higgs-boson mass as the relevant scale.  The leading-order amplitude may be
decomposed into a linear combination of two gauge-invariant subamplitudes,
\begin{eqnarray}
{\cal A}^\mu_0  = {\cal A}_A^\mu + {\cal A}_B^\mu \label{tree}\;.
\end{eqnarray}
These subamplitudes are gauge invariant in the sense that they each satisfy the
Ward identity $p_{3\mu}{\cal A}_{A,B}^\mu =0$, where $p_{3\mu}$ is the gluon
four-momentum.  They are related to the two independent helicity amplitudes for
this subprocess. The explicit form of the subamplitudes and the helicity
amplitudes are given in Appendix~A.

The spin- and color-averaged cross section for $gb\to hb$ is
\begin{equation}
\frac{d\bar\sigma_{gb\to hb}}{dt}= -\frac{1}{s^2}\frac{\alpha_S(\mu)}{24}
\left(\frac{y_b(\mu)}{\sqrt 2}\right)^2 \frac{m_h^4+u^2}{st}\;, \label{treecs}
\end{equation}
where $s,t,u$ are the usual Mandelstam variables (the first and second
diagrams in Fig.~\ref{fig:gbhb} have poles in the $t$ and $s$ channels,
respectively), $\alpha_S(\mu)$ is the $\overline{\rm MS}$ strong coupling, and
$y_b(\mu)$ is the $\overline{\rm MS}$ Yukawa coupling ($y_b(\mu)/\sqrt
2=\overline{m}_b(\mu)/v$ in the standard model, where $\overline{m}_b(\mu)$ is
the $\overline{\rm MS}$ mass, and $v=(\sqrt 2 G_F)^{-1/2} \approx 246$ GeV).
We choose the scale $\mu=m_h$ as our central value. It is important to use
$\overline{m}_b(m_h)$ rather than the pole mass when evaluating the Yukawa
coupling, as the latter is significantly greater than the former, and would
yield an inflated cross section.\footnote{The evaluation of
$\overline{m}_b(m_h)$ is detailed in Ref.~\cite{Dicus:1998hs}. We use
$\overline{m}_b(\overline{m}_b)=4.2$ GeV as the initial condition
\cite{Groom:in}.}  The cross section for the charge-conjugate subprocess
$g\bar b\to h\bar b$ is identical. The cross section is also identical for the
production of a pseudoscalar Higgs boson ($A^0$).

We neglect the $b$-quark mass in Eq.~(\ref{treecs}) and throughout, except in
the evaluation of the Yukawa coupling.  This corresponds to the simplified
ACOT scheme \cite{Aivazis:1993pi,Collins:1998rz,Kramer:2000hn}.  The $b$-quark
mass may be neglected, with no loss of accuracy, in any diagram in which the
$b$ quark is an initial-state parton. Terms proportional to the $b$-quark mass
enter only in the $1/\ln(m_h/m_b)$ correction. This is discussed at the end of
the next section.

\newpage

\section{$1/\ln(m_h/m_b)$ correction}%
\label{sec:invlog}

Consider the subprocess $gg,q\bar q\to b\bar bh$, shown in
Fig.~\ref{fig:ggbbh}. It is of order $\alpha_S^2$ (times the Yukawa
coupling).  Since the leading-order subprocess $gb\to hb$ is of order
$\alpha_S^2\ln(m_h/m_b)$, this subprocess is suppressed by $1/\ln(m_h/m_b)$
relative to the leading-order subprocess (for $m_h \gg m_b$)
\cite{Dicus:1998hs,Stelzer:1997ns}.

The helicity amplitudes for this subprocess are given in Appendix~B.
Integration over the phase space of the final-state particles is divergent
when the $\bar b$ is collinear with an initial gluon,\footnote{This pertains
only to $gg\to b\bar bh$, which makes a much larger contribution than $q\bar
q\to b\bar bh$.} since we use $m_b=0$. This collinear divergence is regulated
using modern dimensional reduction (DR) \cite{Kunszt:1994sd}, and absorbed
into the $b$ distribution function using a dipole-subtraction method
\cite{Ellis:1980wv} as formulated in Ref.~\cite{Catani:1996vz}.\footnote{See
Ref.~\cite{Campbell:2000bg} for details on the implementation of this method.}
This subtraction, together with Gribov-Lipatov-Altarelli-Parisi evolution of
the parton distribution functions, sums terms of order
$\alpha_S^n\ln^n(m_h/m_b)$, to all orders in perturbation theory, into the $b$
distribution function
\cite{Olness:1987ep,Barnett:1987jw,Aivazis:1993pi,Collins:1998rz}.  This
yields a well-behaved perturbation expansion in terms of the parameters
$1/\ln(m_h/m_b)$ and $\alpha_S$ (the latter to be discuss in
Section~\ref{sec:alphas}).  Our final result is in the $\overline{\rm MS}$
factorization scheme.

Some fraction of the events from this subprocess yield a final state with two
$b$ quarks at high $p_T$.  In that case the contribution of this subprocess to
the total cross section is enhanced, since either $b$ can be tagged.  If the
$b$-tagging efficiency is $\epsilon_b$, the probability of tagging one or more
$b$ quarks when both are at high $p_T$ is $2\epsilon_b(1-\epsilon_b) +
\epsilon_b^2$. This results in an enhancement factor of $2-\epsilon_b$
relative to subprocesses in which only one $b$ quark is at high $p_T$. If the
Higgs boson decays to $b\bar b$, the enhancement factor remains $2-\epsilon_b$,
if we demand three or more $b$ tags and also demand that two of these tags
come from the Higgs-boson decay products (so that two $b$-tagged jets
reconstruct the Higgs-boson mass).

Since we neglect the $b$-quark mass throughout the calculation, we are making
an approximation.  To include the $b$-quark mass, one would calculate the
diagrams of Fig.~\ref{fig:ggbbh} with a finite quark mass
\cite{Aivazis:1993pi,Collins:1998rz,Kramer:2000hn}.  This would introduce terms
of order $m_b^2/m_h^2$ and $m_b^2/p_T^2$. Hence the only approximation we are
making by neglecting the $b$-quark mass throughout the calculation is of order
$1/\ln(m_h/m_b)\times m_b^2/m_h^2$ and $1/\ln(m_h/m_b)\times m_b^2/p_T^2$.

\section{$\alpha_S$ correction}
\label{sec:alphas}

In this section we discuss the genuine correction of order $\alpha_S$.  We
divide it into two classes: virtual and real.  Collinear divergences are
isolated and absorbed into the parton distribution functions.  Soft divergences
cancel between the virtual and real corrections.  Both types of divergences are
regulated using modern dimensional reduction (DR) and are cancelled using a
dipole-subtraction method, as in the previous section. Our final result is in
the $\overline{\rm MS}$ factorization scheme.  The $b$-quark mass is neglected
throughout this section; this introduces no approximation
\cite{Aivazis:1993pi,Collins:1998rz,Kramer:2000hn}.

\begin{figure}[p]
\begin{center}
\vspace*{0cm} \hspace*{0cm} \epsfxsize=9cm \epsfbox{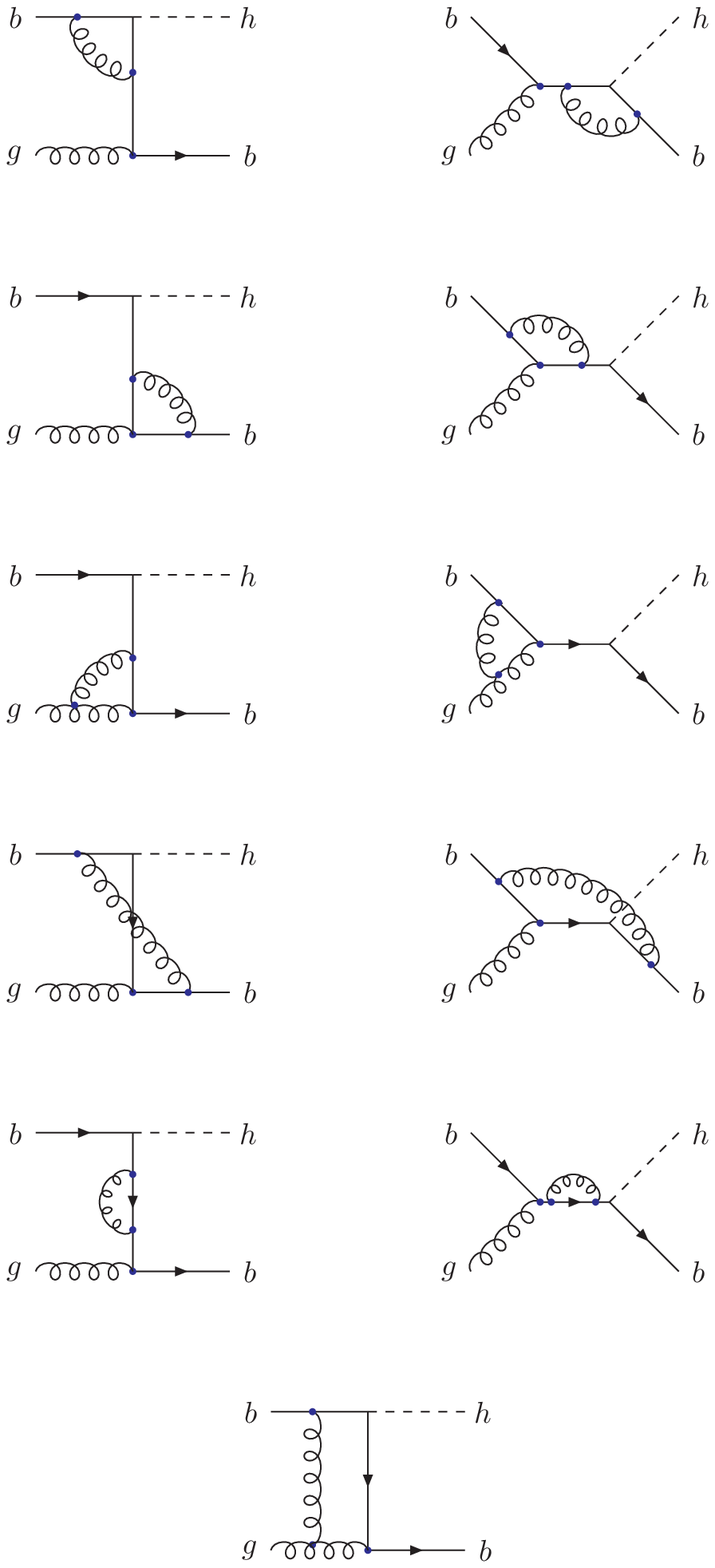} \vspace*{0cm}
\end{center}
\caption{Virtual correction to $gb\to hb$.  External-leg wavefunction
renormalization diagrams (not shown) vanish in modern dimensional reduction
for massless particles.} \label{fig:virtual}
\end{figure}

\subsection{Virtual correction}%
\label{sec:virtual}

The one-loop correction to the subprocess $gb\to hb$ is shown in
Fig.~\ref{fig:virtual}.  We calculate in $d=4-2\epsilon$ dimensions, using
modern dimensional reduction (DR). We also used conventional dimensional
regularization (CDR) as a check on our calculation \cite{Catani:1996pk}, as
discussed in Appendix~E. In DR, the result for the one-loop amplitude is
\begin{eqnarray}
{\cal A}^{\mu}_1 &=& {\cal A}_0^{\mu}\frac{\alpha_S}{4\pi} \left[
\frac{C_A}{2}\left(-2 C(s) + D(s,u) - \frac12 D(s,t) + C'(u) \right) \right.
\nonumber \\
&&\hspace*{2.2cm}\left.+ C_F \left(C(s) - D(s,u) - C'(t) - C'(u) \right)
+ (s \leftrightarrow t)\frac{}{}\right]\nonumber \\
&+&  {\cal A}_B^{\mu} \frac{\alpha_S}{4\pi} (C_A-C_F)\;,
\label{virtual}\end{eqnarray} where the scalar loop integrals $C,C',D$ are
defined in Appendix~C, and $C_F=(N_c^2-1)/2N_c=4/3, C_A=N_c=3$.  The one-loop
amplitude is proportional to the tree amplitude, ${\cal A}_0^\mu$, except for
the last term, which is proportional to one of the two gauge-invariant tree
subamplitudes, ${\cal A}_B^\mu$ [Eq.~(\ref{tree2})], times a finite constant.
We checked that this amplitude has the structure of infrared (soft and
collinear) divergences expected from the dipole-subtraction method (see
Appendix~E).

The above expression contains ultraviolet divergences.  These are cancelled by
the renormalization of the strong and Yukawa couplings, as discussed in
Appendix~E. The ultraviolet divergences are also regulated using modern
dimensional reduction (DR). The renormalization of the Yukawa coupling with
this regulator in the $\overline{\rm MS}$ renormalization scheme is derived in
Appendix~D.

\subsection{Real correction}%
\label{sec:real}

\begin{figure}[tb]
\begin{center}
\vspace*{0cm} \hspace*{0cm} \epsfxsize=10cm \epsfbox{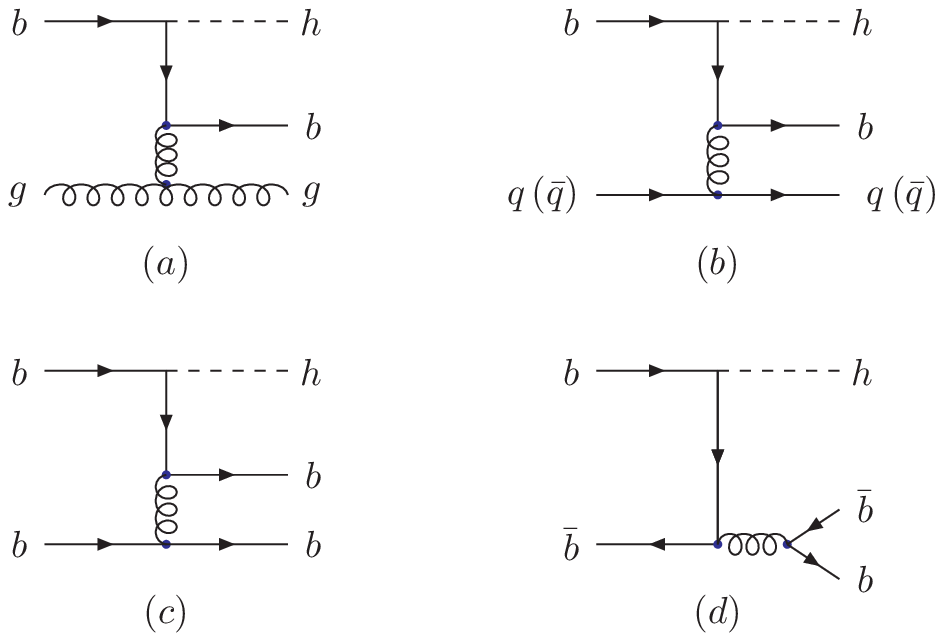} \vspace*{0cm}
\end{center}
\caption{Representative diagrams for subprocesses contributing to the real
correction to $gb\to hb$: (a) $gb\to gbh$ (8~diagrams); (b) $q(\bar q)b\to
q(\bar q)bh$ (2~diagrams); (c) $bb\to bbh$ (8~diagrams); $b\bar b\to b\bar bh$
(8~diagrams).} \label{fig:real}
\end{figure}

The real correction of ${\cal O}(\alpha_S)$ has several contributions.
Fig.~\ref{fig:real}(a) shows the contribution from real gluon emission, $gb\to
gbh$; (b) shows the subprocesses $qb\to qbh$ and $\bar qb\to \bar qbh$; (c)
shows the subprocess $bb\to bbh$; and (d) shows the subprocess $b\bar b\to
b\bar bh$. Another real correction, $gg,q\bar q\to b\bar bh$, shown in
Fig.~\ref{fig:ggbbh}, is of ${\cal O}(1/\ln(m_h/m_b))$; it is discussed in
Section~\ref{sec:invlog}. The helicity amplitudes for these subprocesses are
given in Appendix~B.

The subprocess $bb\to bbh$ (and $b\bar b\to b\bar bh$) requires some additional
consideration.  Since there are two $b$ quarks in the initial state, this
subprocess is of order $\alpha_S^4\ln^2(m_h/m_b)$, which is suppressed relative
to the leading-order subprocess by $\alpha_S^2\ln(m_h/m_b)$.  Thus it is not
truly a correction of order $\alpha_S$.  Nevertheless, it is a
next-to-leading-order correction in powers of $\alpha_S$ and $1/\ln(m_h/m_b)$,
so it is appropriate to include it in our calculation.  Furthermore, this
subprocess yields two $b$ quarks in the final state. Thus, as discussed in
Section~\ref{sec:invlog}, this subprocess is enhanced by a factor
$2-\epsilon_b$ when both $b$ quarks are at high $p_T$.  However, this
contribution is less than one percent of the leading-order cross section, so
this point is moot.

The subprocess $b\bar b\to b\bar bh$ has a contribution from the diagram shown
in Fig.~\ref{fig:real}(d) in which a gluon splits into a final-state $b\bar b$
pair. Since we neglect the $b$ mass throughout our calculation, this subprocess
contains a divergence when the $b$ and $\bar b$ are collinear.  In reality,
the $b$-quark mass regulates this divergence.  To approximate this effect, we
restrict the $b\bar b$ invariant mass to be greater than $2m_b$.  This
correctly captures the dominant, logarithmically-enhanced term of order $\ln
s/m_b^2$.  Since this correction is less than one percent of the leading-order
cross section, this approximation suffices.

\section{Results}%
\label{sec:results}

\begin{figure}[p]
\begin{center}
\vspace*{0cm} \hspace*{0cm} \epsfxsize=14cm \epsfbox{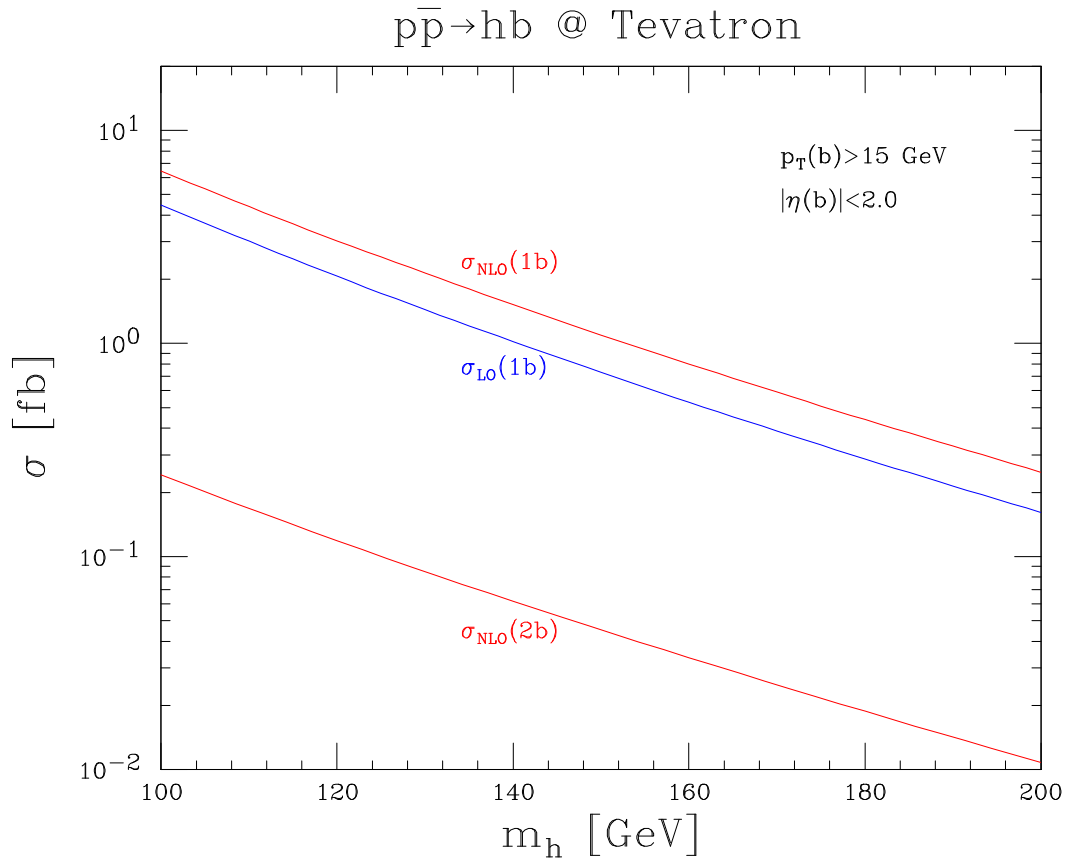} \vspace*{0cm}
\caption{Cross section for the associated production of the Higgs boson and a
single $b$ quark at the Tevatron.  The $b$ quark is within the tagging region
of the silicon vertex detector ($p_T > 15$ GeV, $|\eta|<2$). The curve labeled
$\sigma_{\rm LO}(1b)$ is the leading-order cross section, evaluated with LO
parton distribution functions (CTEQ5L) and couplings evolved at LO, evaluated
at $\mu=m_h$.  The notation indicates that there is only one $b$ quark at high
$p_T$.  The curve labeled $\sigma_{\rm NLO}(1b)$ is the next-to-leading-order
cross section, evaluated with NLO parton distribution functions (CTEQ5M1) and
couplings evolved at NLO, evaluated at $\mu=m_h$.  Only the subprocesses that
yield a single $b$ quark in the tagging region are included.  The cross section
for NLO subprocesses that yield two $b$ quarks in the tagging region is labeled
$\sigma_{\rm NLO}(2b)$.} \label{fig:sigmaTEV}
\end{center}
\end{figure}

\begin{figure}[p]
\begin{center}
\vspace*{0cm} \hspace*{0cm} \epsfxsize=14cm \epsfbox{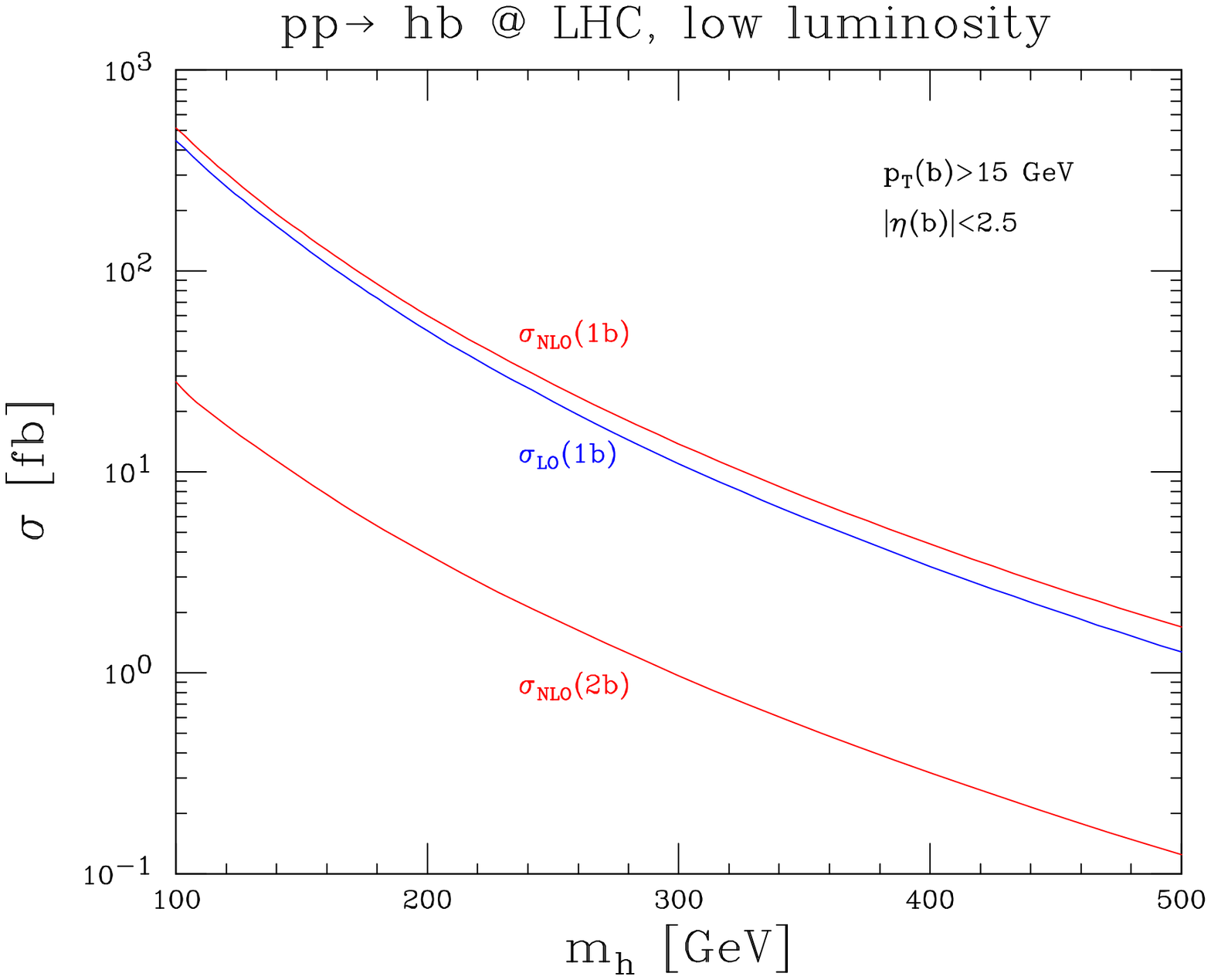}
\vspace*{0cm} \caption{Same as Fig.~\ref{fig:sigmaTEV}, but at the LHC, and
with a $b$-tagging region of $p_T > 15$ GeV, $|\eta|<2.5$.}
\label{fig:sigmaLHC15}
\end{center}
\end{figure}

\begin{figure}[p]
\begin{center}
\vspace*{0cm} \hspace*{0cm} \epsfxsize=14cm \epsfbox{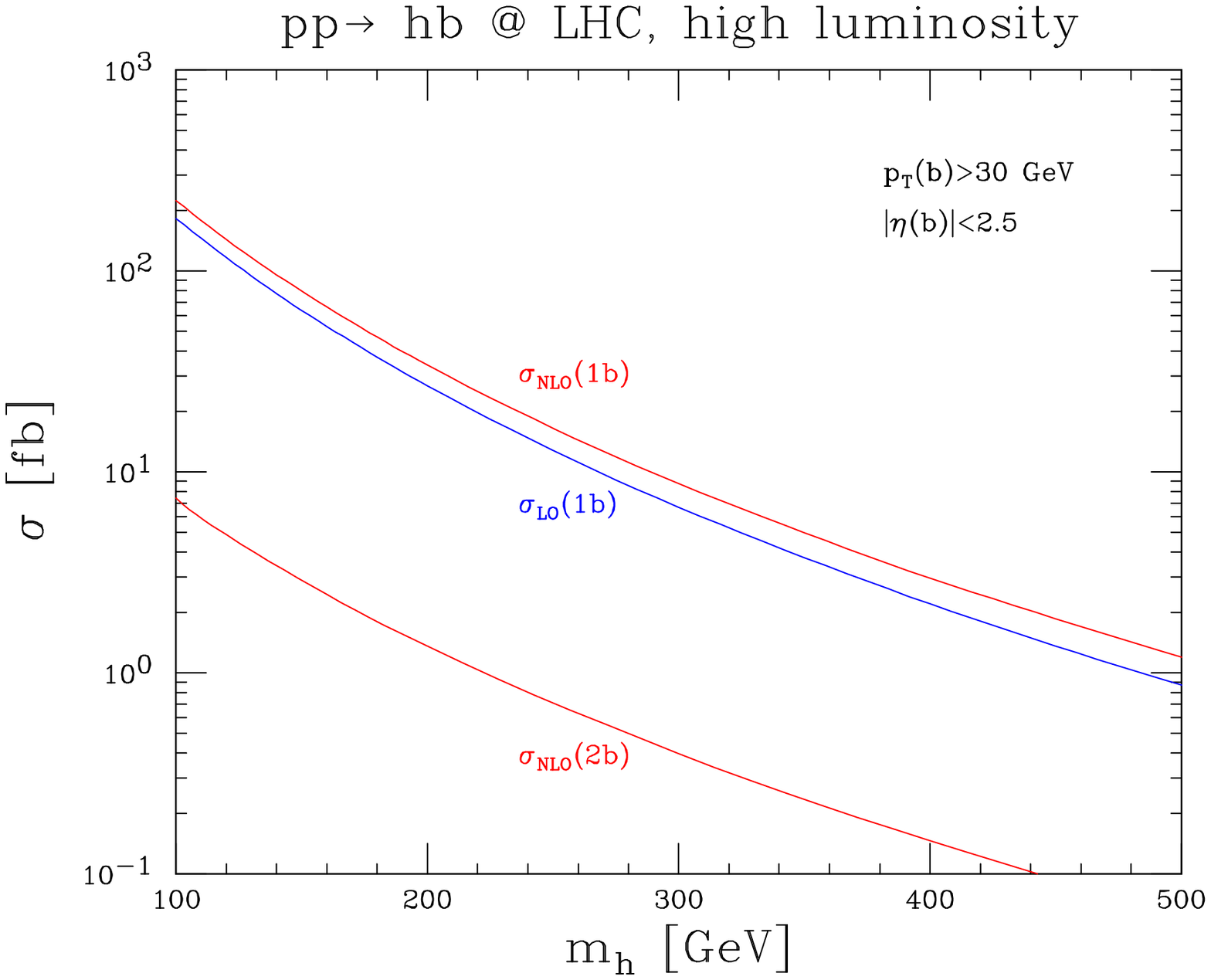}
\vspace*{0cm} \caption{Same as Fig.~\ref{fig:sigmaTEV}, but at the LHC, and
with a $b$-tagging region of $p_T > 30$ GeV, $|\eta|<2.5$.}
\label{fig:sigmaLHC30}
\end{center}
\end{figure}

Figures \ref{fig:sigmaTEV} -- \ref{fig:sigmaLHC30} show the cross sections for
associated production of the Higgs boson and a single bottom quark {\it vs}.\
the Higgs-boson mass at the Tevatron and the LHC.  These cross sections
pertain to both a scalar and a pseudoscalar Higgs boson. The Yukawa coupling
is set to its standard-model value. At the Tevatron, the $b$
jet\footnote{Partons within a cone of $\Delta R = 0.7$ are clustered into a
single $b$ jet.} is required to have a minimum $p_T$ of 15 GeV and a rapidity
of magnitude less than 2, such that it can be tagged by the silicon vertex
detector; we refer to this as the tagging region. At the LHC the rapidity
coverage is taken to be $|\eta(b)|<2.5$. Two plots are given for the LHC, one
with a minimum $p_T$ of 15 GeV (appropriate for low-luminosity running) and
one with 30 GeV (appropriate for high-luminosity running).  Each figure has
three curves.  The curve labeled $\sigma_{\rm LO}(1b)$ is the leading-order
cross section, calculated with LO parton distribution functions (CTEQ5L
\cite{Lai:1999wy}) and couplings evolved at LO, with the factorization and
renormalization scales set to $\mu=m_h$.\footnote{The evolution of
$\alpha_S(\mu)$ uses the value of $\Lambda_{QCD}$ corresponding to the parton
distribution functions.} The notation indicates that there is only one $b$
quark at high $p_T$.  The curve labeled $\sigma_{\rm NLO}(1b)$ is the
next-to-leading-order cross section, calculated with NLO parton distribution
functions (CTEQ5M1) and couplings evolved at NLO, with $\mu=m_h$. Only the
subprocesses that yield a single $b$ quark in the tagging region are included.
Some of the NLO subprocesses yield two $b$ quarks in the tagging region; this
cross section is labeled $\sigma_{\rm NLO}(2b)$ in the figures. This cross
section is dominated by the subprocess $gg\to b\bar bh$, discussed in
Section~\ref{sec:invlog}. The NLO cross section with one or more $b$ tags is
given by $\sigma_{\rm NLO}(1b)\epsilon_b + \sigma_{\rm
NLO}(2b)(2\epsilon_b(1-\epsilon_b)+\epsilon_b^2)$, where $\epsilon_b$ is the
$b$-tagging efficiency.  As is evident from the figures, the NLO cross section
is dominated by the subprocesses with a single $b$ quark in the tagging region.

The NLO correction ranges from $50-60\%$ of the LO cross section at the
Tevatron for $m_h=100-200$ GeV.  At the LHC, the correction ranges from
$20-40\%$ for $p_T>15$ GeV, and $25-45\%$ for $p_T>30$ GeV, for $m_h=120-500$
GeV.\footnote{This is the size of the correction for $\mu=m_h$.  The
correction is less for smaller values of $\mu$.} Most of the correction comes
from the ${\cal O}(\alpha_S)$ contribution. The ${\cal O}(1/\ln(m_h/m_b))$
contribution is small, less than $10\%$ of the LO cross section.  Thus the
terms we are neglecting by using $m_b=0$ throughout the calculation, of order
$1/\ln(m_h/m_b)\times m_b^2/m_h^2$ and $1/\ln(m_h/m_b)\times m_b^2/p_T^2$, are
very small.

As discussed in the Introduction, recent analyses for the decay $h\to b\bar b$
use $gg,q\bar q\to hb\bar b$ as the Higgs-boson production subprocess, and
demand a final state with four jets, with either at least three $b$ tags, or
with four $b$ tags \cite{unknown:1999fr,Carena:1998gk,Richter-Was:1997gi,
Diaz-Cruz:1998qc,Balazs:1998nt,Carena:2000yx}. The cross section with at least
three $b$ tags (two of which come from the decay products of the Higgs
boson)\footnote{A factor $BR(h\to b\bar b)\epsilon_b^2$ is implicit in the
following cross sections.} is $\sigma_{\rm
NLO}(2b)(2\epsilon_b(1-\epsilon_b)+\epsilon_b^2)$; with four $b$ tags it is
$\sigma_{\rm NLO}(2b)\epsilon_b^2$.  Both of these are an order of magnitude
less than the cross section with three or more jets, with three or more $b$
tags, given by $\sigma_{\rm NLO}(1b)\epsilon_b + \sigma_{\rm
NLO}(2b)(2\epsilon_b(1-\epsilon_b)+\epsilon_b^2)$. Thus our motivation for
carrying out this calculation was well founded.

%%%%%%%%%%%%%%%%%%%%%%%%%%%%%%%%%%%%%%%%%%%%%%%%%%%%%%%%%%%%%%%%%%%%%%%%%%%
\begin{table}[t]
\caption{Cross sections (fb) for the associated production of the Higgs boson
and a single $b$ quark at the Tevatron.  The central value corresponds to the
choice of factorization and renormalization scale $\mu=m_h$; these values are
plotted in Fig.~\ref{fig:sigmaTEV}.  The uncertainty corresponds to varying
the scale from $\mu=m_h/2$ to $\mu=2m_h$.  The $b$ quark is within the tagging
region of the silicon vertex detector ($p_T > 15$ GeV, $|\eta|<2$). The column
labeled $\sigma_{\rm LO}(1b)$ is the leading-order cross section, evaluated
with LO parton distribution functions (CTEQ5L) and couplings evolved at LO.
The notation indicates that there is only one $b$ quark at high $p_T$. The
column labeled $\sigma_{\rm NLO}(1b)$ is the next-to-leading-order cross
section, evaluated with NLO parton distribution functions (CTEQ5M1) and
couplings evolved at NLO.  Only the subprocesses that yield a single $b$ quark
in the tagging region are included.  The cross section for NLO subprocesses
that yield two $b$ quarks in the tagging region is labeled $\sigma_{\rm
NLO}(2b)$.} \addtolength{\arraycolsep}{0.1cm}
\renewcommand{\arraystretch}{1.2}
\medskip
\begin{center}
\begin{tabular}[4]{|c|ccc|}
\hline \hline
& \multicolumn{3}{c|}{$p\bar{p}$ @ $\sqrt{s}=2$ TeV  }\\[1pt]
\cline{2-4}
$m_h$ (GeV) & \multicolumn{3}{c|}{$p_T(b)>15$ GeV}\\
%\cline{2-4}
            & $\sigma_{\rm LO}(1b)$ & $\sigma_{\rm NLO}(1b)$  & $\sigma_{\rm NLO}(2b)$  \\
\hline
$100$ &$  4.49    ^{+19\%}_{-17\%}      $&$ 6.45^{+0\%}_{-4\%} $ & $  0.24    ^{+62\%}_{-35\%} $\\[7pt]
$120$ &$  2.06    ^{+22\%}_{-18\%}      $&$ 3.03^{+2\%}_{-5\%} $ & $  0.12    ^{+62\%}_{-35\%} $\\[7pt]
$140$ &$  1.02    ^{+23\%}_{-19\%}      $&$ 1.52^{+3\%}_{-6\%} $ & $  0.062   ^{+62\%}_{-35\%} $\\[7pt]
$160$ &$  0.529   ^{+25\%}_{-19\%}      $&$ 0.80^{+2\%}_{-8\%} $ & $  0.034   ^{+63\%}_{-35\%} $\\[7pt]
$180$ &$  0.287   ^{+26\%}_{-20\%}      $&$ 0.44^{+3\%}_{-8\%} $ & $  0.019   ^{+63\%}_{-36\%} $\\[7pt]
$200$ &$  0.162   ^{+27\%}_{-21\%}      $&$ 0.25^{+4\%}_{-8\%} $ & $  0.011   ^{+63\%}_{-36\%} $\\[7pt]
\hline \hline
\end{tabular}
\end{center}
\label{tab:TEV}
\end{table}
%%%%%%%%%%%%%%%%%%%%%%%%%%%%%%%%%%%%%%%%%%%%%%%%%%%%%%%%%%%%%%%%%%%%%%%%%%%

%%%%%%%%%%%%%%%%%%%%%%%%%%%%%%%%%%%%%%%%%%%%%%%%%%%%%%%%%%%%%%%%%%%%%%%%%%%
\begin{table}[t]
\caption{Same as Table~\ref{tab:TEV}, but at the LHC.  The left side of the
table corresponds to a $b$-tagging region of $p_T > 15$ GeV, $|\eta|<2.5$,
appropriate for low-luminosity running.  These cross sections are plotted in
Fig.~\ref{fig:sigmaLHC15}.  The right side of the table corresponds to $p_T >
30$ GeV, $|\eta|<2.5$, appropriate for high-luminosity running.  These cross
sections are plotted in Fig.~\ref{fig:sigmaLHC30}.}
\addtolength{\arraycolsep}{0.1cm}
\renewcommand{\arraystretch}{1.2}
\medskip
\begin{center}
\begin{tabular}[4]{|c|ccc|ccc|}
\hline \hline
& \multicolumn{6}{c|}{$pp$ @ $\sqrt{s}=14$ TeV }\\[3pt]
\cline{2-7} $m_h$ (GeV) & \multicolumn{3}{c|}{$p_T(b)>15$ GeV} &
\multicolumn{3}{c|}{$p_T(b)>30$ GeV}
\\[1pt]
& $\sigma_{\rm LO}(1b)$ & $\sigma_{\rm NLO}(1b)$  & $\sigma_{\rm NLO}(2b)$  &
$\sigma_{\rm LO}(1b)$ & $\sigma_{\rm NLO}(1b)$  &
$\sigma_{\rm NLO}(2b)$ \\[2pt]
\hline
$120$ &$  269    ^{+5 \%}_{-9 \%} $&$ 305 ^{-1\%}_{+1\%}  $&$ 17     ^{+38\%}_{-25\%}  $&$  117    ^{+7 \%}_{-9 \%} $&$143  ^{+1\%}_{-1\%} $ &$ 4.9  ^{+40 \%}_{-26\%}$ \\[7pt]
$160$ &$  108    ^{+10\%}_{-10\%} $&$ 127 ^{-2\%}_{+0\%}  $&$ 7.7    ^{+37\%}_{-25\%}  $&$  52.8   ^{+10\%}_{-10\%} $&$66.2 ^{+1\%}_{-3\%} $ &$ 2.5  ^{+40 \%}_{-26\%}$ \\[7pt]
$200$ &$  49.9   ^{+13\%}_{-13\%} $&$ 60.1^{-1\%}_{-1\%}  $&$ 3.9    ^{+39\%}_{-25\%}  $&$  26.8   ^{+12\%}_{-12\%} $&$34.0 ^{+1\%}_{-3\%} $ &$ 1.4  ^{+40 \%}_{-27\%}$ \\[7pt]
$300$ &$  11.0   ^{+15\%}_{-12\%} $&$ 13.8^{-1\%}_{-2\%}  $&$ 1.0    ^{+40\%}_{-26\%}  $&$  6.67   ^{+14\%}_{-13\%} $&$8.8  ^{+2\%}_{-4\%} $ &$ 0.40 ^{+41 \%}_{-27\%}$ \\[7pt]
$400$ &$  3.39   ^{+16\%}_{-14\%} $&$ 4.37^{+0\%}_{-3\%}  $&$ 0.32   ^{+40\%}_{-26\%}  $&$  2.21   ^{+16\%}_{-14\%} $&$2.96 ^{+2\%}_{-4\%} $ &$ 0.15 ^{+41 \%}_{-27\%}$ \\[7pt]
$500$ &$  1.27   ^{+18\%}_{-15\%} $&$ 1.69^{+0\%}_{-4\%}  $&$ 0.12   ^{+42\%}_{-27\%}  $&$  0.872  ^{+18\%}_{-15\%} $&$1.20 ^{+2\%}_{-5\%} $ &$ 0.062^{+41 \%}_{-28\%}$ \\[7pt]
\hline \hline
\end{tabular}
\end{center}
\label{tab:LHC}
\end{table}
%%%%%%%%%%%%%%%%%%%%%%%%%%%%%%%%%%%%%%%%%%%%%%%%%%%%%%%%%%%%%%%%%%%%%%%%%%%

Similarly, the existing studies of $h\to \tau^+\tau^-,\mu^+\mu^-$ with at
least one $b$ tag use $gg,q\bar q\to b\bar bh$ as the Higgs-boson production
subprocess \cite{unknown:1999fr,Drees:1997sh,Carena:1998gk}.  One should
instead use the NLO calculation of $gb\to hb$, since this is a much larger
cross section.

\begin{figure}[p]
\begin{center}
\vspace*{0cm} \hspace*{0cm} \epsfxsize=14cm \epsfbox{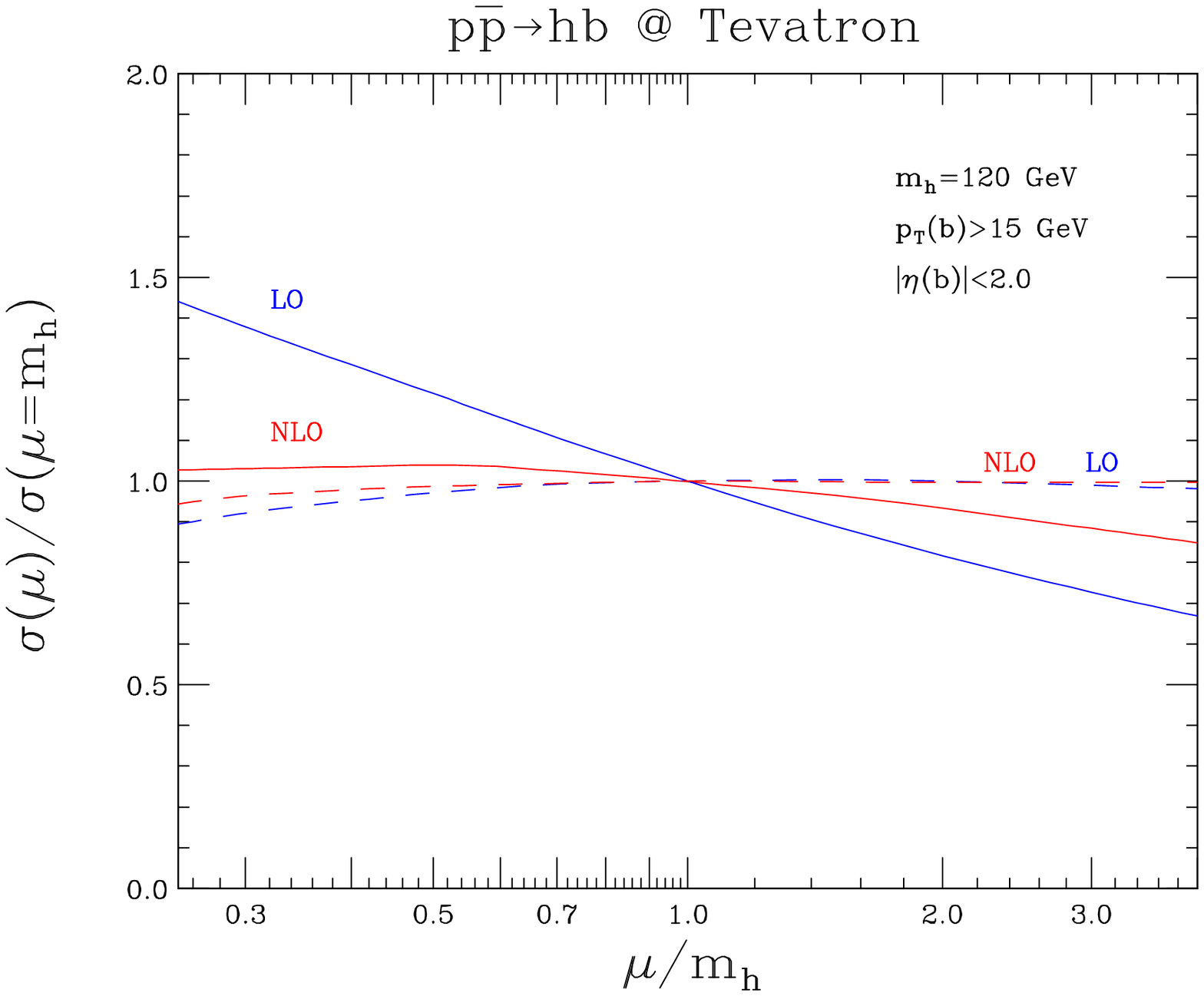} \vspace*{0cm}
\caption{Cross section for the associated production of the Higgs boson and a
single $b$ quark {\it vs}.\ the common factorization and renormalization scale
$\mu$, for $m_h=120$ GeV at the Tevatron (solid curves).  The ratio of the
cross section at scale $\mu$ to the cross section at scale $\mu=m_h$ is
plotted {\it vs}.\ the ratio of the scales.  The next-to-leading-order (NLO)
cross section is less sensitive to the scale $\mu$ than the leading-order (LO)
cross section.  Also shown is the dependence on the factorization scale alone,
with the renormalization scale fixed at $\mu=m_h$ (dashed curves).}
\label{fig:mu}
\end{center}
\end{figure}

\begin{figure}[p]
\begin{center}
\vspace*{0cm} \hspace*{0cm} \epsfxsize=14cm \epsfbox{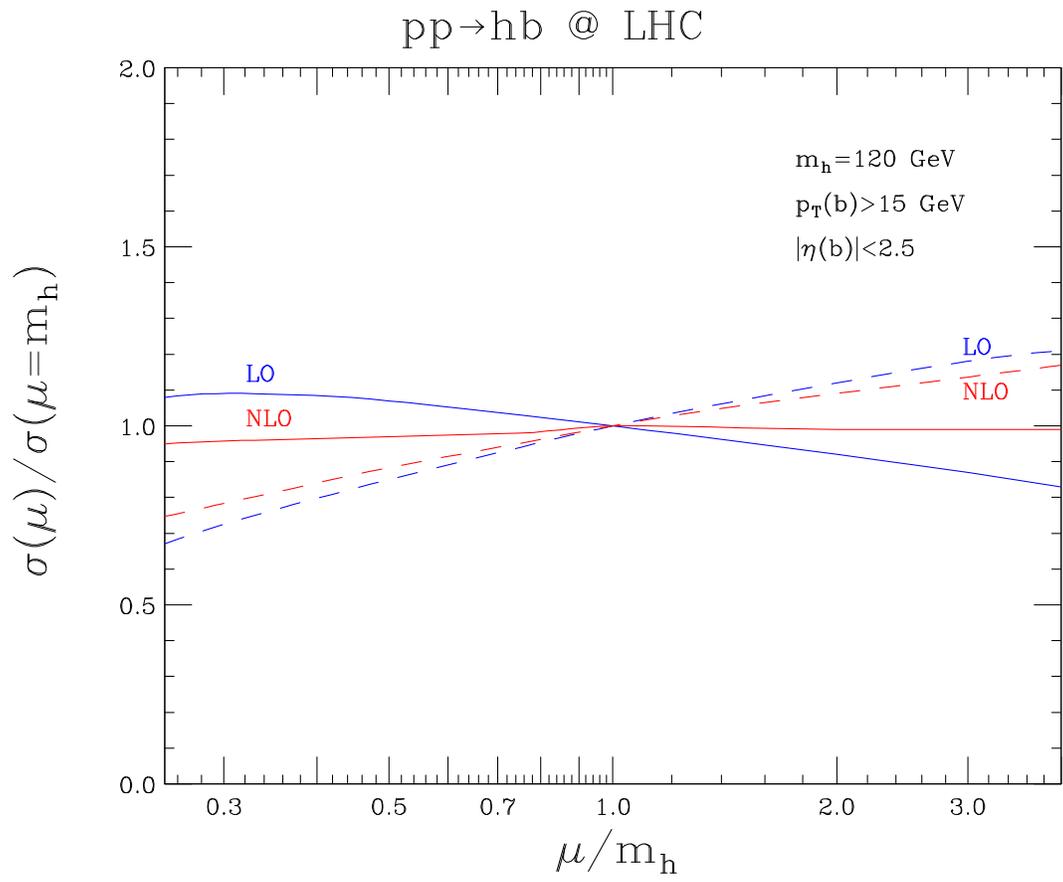}
\vspace*{0cm} \caption{Same as Fig.~\ref{fig:mu}, but at the LHC.}
\label{fig:mulhc}
\end{center}
\end{figure}

The NLO calculation of the cross section for associated production of the
Higgs boson and a single $b$ quark gives a more accurate estimate of the cross
section than the LO calculation.  This is evidenced by the fact that the NLO
calculation of the cross section is less sensitive to the choice of
factorization and renormalization scales than the LO calculation.  Typical
examples are shown in Figs.~\ref{fig:mu} and \ref{fig:mulhc}, where we plot
the LO and NLO cross section {\it vs}.\ the common factorization and
renormalization scale $\mu$, for $m_h=120$ GeV at the Tevatron and the LHC
(solid curves) .\footnote{In this and the following figure, we take
$\epsilon_b=1$ when combining $\sigma_{\rm NLO}(1b)$ and $\sigma_{\rm
NLO}(2b)$ to obtain the NLO cross section. However, the NLO cross section is
dominated by $\sigma_{\rm NLO}(1b)$ for any value of $\epsilon_b$.} In Tables
\ref{tab:TEV} and \ref{tab:LHC} we give the cross section evaluated at
$\mu=m_h$ as the central value (these are the numbers plotted in
Figs.~\ref{fig:sigmaTEV} -- \ref{fig:sigmaLHC30}), with uncertainties
corresponding to $\mu = m_h/2$ (upper uncertainty) and $\mu=2m_h$ (lower
uncertainty).  The scale dependence is significantly reduced when going from
LO to NLO. Our NLO cross section can be used to normalize any future studies
that make use of this production mechanism.

Also shown in Figs.~\ref{fig:mu} and \ref{fig:mulhc} is the factorization-scale
dependence of the cross section, with the renormalization scale fixed to
$\mu=m_h$ (dashed curves).  The factorization-scale dependence decreases at
NLO, as expected.  At the Tevatron, the factorization-scale dependence is
negligible, even at LO.  At the LHC, the factorization-scale dependence is
greater than the dependence on the common factorization and renormalization
scales.  This indicates that there is compensation between the factorization
and renormalization scales when the two are varied simultaneously.

\begin{figure}[p]
\begin{center}
\vspace*{0cm} \hspace*{0cm} \epsfxsize=14cm \epsfbox{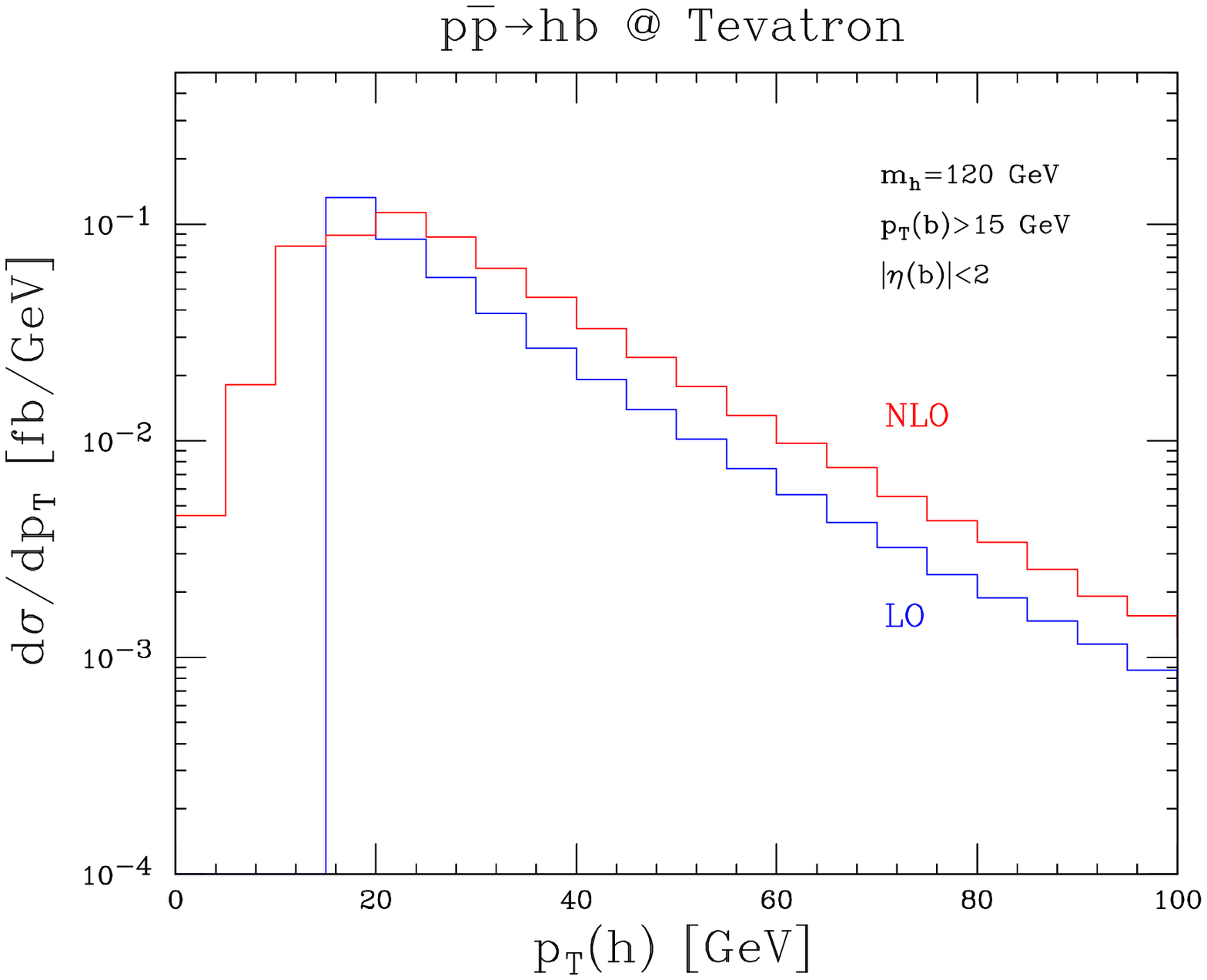} \vspace*{0cm}
\caption{Higgs-boson $p_T$ distribution for associated production of the Higgs
boson and a single $b$ quark, for $m_h=120$ GeV at the Tevatron. At leading
order (LO) the Higgs-boson $p_T$ is balanced against that of the $b$ quark,
while at next-to-leading-order (NLO) it is balanced against that of the $b$
quark and an additional parton.} \label{fig:pt}
\end{center}
\end{figure}

The uncertainty in the choice of factorization and renormalization scales
yields some uncertainty in the NLO cross section.  In addition, there is an
uncertainty in the cross section of about $10\%$ due to the uncertainty in the
Yukawa coupling ($\overline{m}_b(\overline{m}_b) = 4.2 \pm 0.2$), and of about
$4\%$ due to the uncertainty in the strong coupling \cite{Groom:in}. The
uncertainty in the gluon distribution function (which also reflects itself in
the uncertainty in the $b$ distribution function) is the source of another
$10\%$ uncertainty in the cross section \cite{Huston:1998jj}.

Recall that it is only valid to use $gg,q\bar q\to b\bar bh$ as the production
subprocess when both $b$ quarks are at high $p_T$.  To demonstrate this, we
evaluated the cross section for the production of the Higgs boson and one
high-$p_T$ $b$ quark using this subprocess by integrating over the momentum of
the other $b$ quark.  For $m_h=120$ GeV, this underestimates the NLO cross
section by a factor of 4.6 at the Tevatron and 2.7 at the LHC.  This factor is
even larger for heavier Higgs bosons.

We also studied the kinematics of the Higgs boson at NLO {\it vs}.\ LO. The
rapidity distribution of the Higgs boson remains almost unchanged.  The $p_T$
distribution of the Higgs boson does change at low $p_T$, as shown in
Fig.~\ref{fig:pt}.  At LO, the $p_T$ of the Higgs boson is balanced against
that of the $b$ quark, so the Higgs-boson $p_T$ cannot be less than the
minimum $p_T$ of the $b$ quark. This restriction is lifted at NLO, since the
$p_T$ of the Higgs boson is balanced against that of the $b$ quark and an
additional parton.

\section{Conclusions}%
\label{sec:conclusions}

Previous studies of the associated production of the Higgs boson and a
high-$p_T$ bottom quark have used $gg,q\bar q\to b\bar bh$ as the production
mechanism
\cite{Stange:ya,Dicus:1988cx,Kunszt:1991qe,unknown:1999fr,Drees:1997sh,Carena:1998gk,
Dai:1994vu,Dai:1996rn,Richter-Was:1997gi,Diaz-Cruz:1998qc,
Balazs:1998nt,Carena:2000yx}, which is valid only if both $b$ quarks are at
high $p_T$.  In this paper we have shown that the cross section for $gb\to hb$
\cite{Choudhury:1998kr,Huang:1998vu} is an order of magnitude larger than that
of $gg,q\bar q\to b\bar bh$. This production mechanism improves the prospects
for the discovery of a Higgs boson with enhanced coupling to the $b$ quark. We
evaluated the cross section for this subprocess at the Tevatron and the LHC at
next-to-leading order in QCD. These cross sections can be used to normalize
any future studies of this production mechanism.  They pertain to both a
scalar and a pseudoscalar Higgs boson.  We have included $gb\to hb$ in the
multi-purpose NLO Monte Carlo program MCFM
\cite{Campbell:2000bg,Ellis:1999ec}. We encourage studies of the signal and
backgrounds for associated production of the Higgs boson with a single high
$p_T$ bottom quark.

\section*{Acknowledgments}

\indent\indent We are grateful for conversations and correspondence with
C.~Oleari and Z.~Tr\'ocs\'anyi.  This work was supported in part by the
U.~S.~Department of Energy under contracts Nos.~DE-AC02-76CH03000 and
DE-FG02-91ER40677.

\section*{Appendix~A}%
\label{sec:appa}

\begin{figure}[hb]
\begin{center}
\vspace*{0cm} \hspace*{0cm} \epsfxsize=5cm \epsfbox{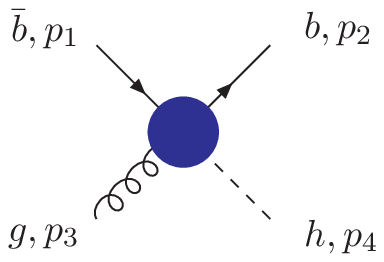} \vspace*{0cm}
\end{center}
\caption{Four-momenta (all outgoing) of the particles for the (unphysical)
subprocess $0\to \bar bbgh$.  The arrows indicate the flow of fermion number.}
\label{fig:0bbgh}
\end{figure}

In this Appendix~we present explicit results for the leading-order subprocess
$gb\to hb$.  In order to be systematic, we give results for the (unphysical)
crossed subprocess $0\to \bar bbgh$, in which all particles are taken to be
outgoing, as shown in Fig.~\ref{fig:0bbgh}. The amplitudes for the physical
subprocesses $gb\to hb$ and $g\bar b\to h\bar b$ may then be obtained by
crossing.  The $b$-quark mass is neglected throughout.  All expressions are
presented in $d=4-2\epsilon$ dimensions, using modern dimensional reduction.

The amplitude for the leading-order subprocess $0\to \bar bbgh$ may be written
in terms of the four-momenta of the $b$, $\bar b$, and gluon.  It is a linear
combination of two gauge-invariant subamplitudes,
\begin{eqnarray}
{\cal A}_0& \equiv & \epsilon_\mu ^{h_3*}(p_3) {\cal A}^\mu_0  = \epsilon_\mu
^{h_3*}(p_3)
({\cal A}_A^\mu + {\cal A}_B^\mu) \label{treeapp}\\
{\cal A}_A^\mu &=&  i \mu^{2\epsilon} g_S \frac{y_b}{\sqrt 2} \sqrt{2} T^a 2
\langle 2^{h_2} | 1^{-h_2} \rangle \left( \frac{p_2^\mu}{s_{23}} -
\frac{p_1^\mu}{s_{13}} \right)
\label{tree1}\\
{\cal A}_B^\mu &=&  i \mu^{2\epsilon} g_S \frac{y_b}{\sqrt 2}
 \sqrt{2} T^a
 \frac{m_h^2-s_{12}}{s_{23} s_{13}}
\langle 2^{h_2}| \gamma^\mu  \hat{p}_3  | {1^{-h_2}}  \rangle\;,\label{tree2}
\end{eqnarray}
where $y_b$ is the $\overline{\rm MS}$ Yukawa coupling ($y_b(\mu)/\sqrt
2=\overline{m}_b(\mu)/v$ in the standard model, where $\overline{m}(\mu)$ is
the $\overline{\rm MS}$ mass and $v=(\sqrt 2 G_F)^{-1/2} \approx 246$ GeV),
$T^a$ are the fundamental-representation matrices of $SU(3)$ (Tr $T^aT^b=
\delta^{ab}/2$), $s_{ij} \equiv (p_i+p_j)^2$, $\hat{p}_3 \equiv \gamma_\mu
p_3^\mu$, $| {1^{\pm}} \rangle \equiv v^{\mp}(p_1)$, $\langle 2^\pm | \equiv
\overline {u^{\pm}}(p_2)$, $\mu$ is the 't Hooft mass (introduced such that
the renormalized couplings are dimensionless in $d$ dimensions), $h_2$ denotes
the helicity of the $b$ quark (the $\bar b$ has the same helicity in the
massless approximation), and $h_3$ denotes the helicity of the gluon. The
subamplitudes are gauge invariant in the sense that they each satisfy the Ward
identity $p_{3\mu}{\cal A}_{A,B}^\mu =0$.

One may also describe this subprocess in terms of helicity amplitudes. We
define helicity amplitudes, $A$, with the overall factors of the coupling
constants removed,
\begin{eqnarray}
{\cal A}_0(1^{h_1}_{\bar{b}},2^{h_2}_b,3^{h_3}_g)&=& \mu^{2\epsilon} g_S
\frac{y_b}{\sqrt 2} \sqrt{2} T^a A(1^{h_1}_{\bar{b}},2^{h_2}_b,3^{h_3}_g)\;,
\label{treehel}
\end{eqnarray}
where the Higgs-boson four-momentum is tacit. There are two independent
helicity amplitudes,
\begin{eqnarray}
A(1_{\bar{b}}^+,2_b^+,3_g^+)&=&  i \frac{m_h^2}{\langle 13\rangle
\langle 23\rangle} \\
A(1_{\bar{b}}^+,2_b^+,3_g^-)&=&  i \frac{[12]^2}{[13][23]}\;,
\end{eqnarray}
using the spinor inner-product notation as reviewed in
Ref.~\cite{Mangano:1990by}.  The helicity-reversed amplitudes are equal to
these (times $-1$ if the Higgs boson is a pseudoscalar), by parity.  The
helicity amplitudes are related to the amplitudes ${\cal A}_A^\mu$ and ${\cal
A}_B^\mu$ above by \begin{eqnarray} &&\epsilon^{\pm *}_\mu(p_3) {\cal
A}_A^\mu= {\cal A}_0(1_{\bar{b}}^\mp,
2_b^\mp,3_g^\pm)\\
&&\epsilon^{\pm *}_\mu(p_3) {\cal A}_B^\mu= {\cal
A}_0(1_{\bar{b}}^\pm,2_b^\pm,3_g^\pm)\;.
\end{eqnarray}
Squaring the amplitude and summing over colors and helicities gives
\begin{equation}
\sum_{\rm col,hel} |{\cal A}_0|^2 = {\rm sgn}(s_{12})  16 \mu^{4\epsilon} g_S^2
\left(\frac{y_b}{\sqrt 2}\right)^2 \frac{m_h^4+s_{12}^2}{s_{13} s_{23}}\;.
\end{equation}
The spin- and color-averaged cross section for the physical subprocess $gb\to
hb$, Eq.~(\ref{treecs}), is then obtained from the above by crossing ($s_{12}
\to u$, $s_{13} \to s$, $s_{23} \to t$).

In the dipole-subtraction method, the tensor
\begin{eqnarray}
&&\sum_{{\rm col},h_{2}=\pm 1}  {\cal A}_0^{\mu}{\cal A}_0^{\nu*} = 8
\mu^{4\epsilon} g_S^2 \left(\frac{y_b}{\sqrt 2}\right)^2 \left[ -g^{\mu\nu}
\frac{(s_{13}+s_{23})^2}{ s_{13} s_{23}}  +4 m_h^2 \left(
\frac{p_2^\mu}{s_{23}} - \frac{p_1^\mu}{s_{13}} \right) \left(
\frac{p_2^\nu}{s_{23}} - \frac{p_1^\nu}{s_{13}} \right) \right.-
\nonumber\\
&& \left. 4\frac{ s_{12} }{ s_{13} s_{23}} p_3^\mu p_3^\nu -2 \frac{ s_{12}
s_{13} -m_h^2 s_{23}}{ s_{13}^2 s_{23}} ( p_3^\mu p_1^\nu + p_3^\nu p_1^\mu) -2
\frac{ s_{12} s_{23} -m_h^2 s_{13}}{ s_{13} s_{23}^2} ( p_3^\mu p_2^\nu +
p_3^\nu p_2^\mu) \right]
\end{eqnarray}
is also needed.

\section*{Appendix~B}%
\label{sec:appb}

We present the helicity amplitudes for the $2\to 3$ subprocesses shown in
Figs.~\ref{fig:ggbbh} and \ref{fig:real}.  All amplitudes are calculated in
$d=4$ dimensions. The amplitudes in $d=4-2\epsilon$ dimensions using modern
dimensional reduction may be obtained via $g_S \to g_S\mu^\epsilon$, $y_b \to
y_b\mu^\epsilon$. The calculations were checked with the code MADGRAPH
\cite{Stelzer:1994ta}.

The subprocesses shown in Figs.~\ref{fig:ggbbh}(a) and \ref{fig:real}(a) may be
obtained from the helicity amplitudes for the unphysical subprocess $0 \to
{\bar b} b g g h$. These helicity amplitudes may be written as
\begin{equation}
{\cal A}(1_{\bar{b}}^{h_1},2_b^{h_2},3_g^{h_3},4_g^{h_4})= (g_S \sqrt{2})^2
\frac{y_b}{\sqrt 2} \left( \frac{\{T^a,T^b\}}{2} A_s + \frac{[T^a,T^b]}{2} A_a
\right)\;,
\end{equation}
where $A_s$ and $A_a$ are the symmetric and antisymmetric combination of
color-ordered amplitudes. The three independent helicity configurations are
given by ($s_{ijk}\equiv (p_i+p_j+p_k)^2$)
\begin{eqnarray}
&&A_s(1_{\bar{b}}^+,2_b^+;3_g^+,4_g^+)=-i \frac{m_h^2\br(1,2)}
  {\br(1,3)\br(1,4)\br(2,3)\br(2,4)}\\
&&A_a(1_{\bar{b}}^+,2_b^+;3_g^+,4_g^+)= i \frac{m_h^2}{\br(3,4)}
\left(\frac{1}{\br(1,3)\br(2,4)}+\frac{1}{
       \br(1,4) \br(2,3)}  \right)
 \\
&&A_s(1_{\bar{b}}^+,2_b^+;3_g^-,4_g^-)=i \frac{{\sq(1,2)}^3}
  {\sq(1,3)\sq(1,4)
    \sq(2,3)\sq(2,4)}
 \\
&&A_a(1_{\bar{b}}^+,2_b^+;3_g^-,4_g^-)=-i \frac{{\sq(1,2)}^2}{\sq(3,4)}
\left(\frac{1}{\sq(1,3)\sq(2,4)}+\frac{1}{
       \sq(1,4) \sq(2,3)}  \right)
\\
&&A_s(1_{\bar{b}}^+,2_b^+;3_g^+,4_g^-)= i \left( \frac{\sq(1,2)\t(1,2,3)}
   {\br(1,3)\br(2,3)
     \sq(1,4)\sq(2,4)} \right. \nonumber \\
     &&+\left.
  \frac{\sq(1,3)
     \left( \br(1,4)\sq(1,2) -
       \br(3,4)\sq(2,3) \right) }
     {\br(1,3)\sq(1,4)\t(1,3,4)} +
  \frac{\left( \br(2,4)
        \sq(1,2) +
       \br(3,4)\sq(1,3) \right)
     \sq(2,3)}{\br(2,3)
     \sq(2,4)\t(2,3,4)} \right)
 \\
&&A_a(1_{\bar{b}}^+,2_b^+;3_g^+,4_g^-)= i  \left[\frac{
  \br(3,4)\sq(2,3)-\br(1,4)\sq(1,2)}
{\br(1,3)\s(3,4) } \left( - \frac{\sq(2,3)} {\sq(2,4)}   +
 \frac{\left( -\s(1,3) + \s(1,4) \right)
 \sq(1,3)}{\sq(1,4) \t(1,3,4)} \right) \right. \nonumber \\
              && - \left.
 \frac{ \br(2,4)   \sq(1,2) +  \br(3,4)\sq(1,3)}{\br(2,3) \s(3,4)}
  \left( \frac{\sq(1,3)}
            {\sq(1,4)} +
           \frac{\left( \s(2,3) - \s(2,4) \right)
              \sq(2,3)}{\sq(2,4)
              \t(2,3,4)} \right)\right]\;.
\end{eqnarray}
Since these are not color-ordered amplitudes, the order of the gluons in the
arguments of the functions $A_s$ and $A_a$ is irrelevant.  Squaring and summing
over colors gives
\begin{equation} \sum_{\rm col}|{\cal
A}(1_{\bar{b}}^{h_1},2_b^{h_2},3_g^{h_3},4_g^{h_4})|^2=
 g_S^4 \left(\frac{y_b}{\sqrt 2}\right)^2  \frac{N_c^2-1}{2}
\left( \frac{N_c^2-2}{N_c} |A_s|^2 + N_c   |A_a|^2 \right)\;.
\end{equation}

The four-quark subprocess of Figs.~\ref{fig:ggbbh}(b) and \ref{fig:real}(b) may
be obtained from the helicity amplitudes for the unphysical subprocess $0 \to
{\bar b} b {\bar q} q h$,
\begin{equation}
{\cal A}(1_{\bar{b}}^{h_1},2_b^{h_2},3_{\bar q}^{h_3},4_q^{h_4})= (g_S
\sqrt{2})^2  \frac{y_b}{\sqrt 2} \frac12 \left(
\delta_{i_4}^{\bar{\imath}_1}\delta_{i_2}^{\bar{\imath}_3} - \frac{1}{N_c}
\delta_{i_2}^{\bar{\imath}_1}\delta_{i_4}^{\bar{\imath}_3} \right) A\;.
\end{equation}
The two independent helicity configurations are given by
\begin{eqnarray}
&&A(1_{\bar{b}}^+,2_b^+,3_{\bar
q}^+,4_q^-)=f(1,2,3,4)+f(2,1,3,4)\\
&&A(1_{\bar{b}}^+,2_b^+,3_{\bar q}^-,4_q^+)=f(1,2,4,3)+f(2,1,4,3)\;,
\end{eqnarray}
where
\begin{equation}
f(1,2,3,4)=i \frac{\,\sq(1,3)\,
     \left( \br(1,4)\,\sq(1,2) -
       \br(3,4)\,\sq(2,3) \right) }
     {\s(3,4) \t(1,3,4)}\;.
\end{equation}
Squaring and summing over colors gives
\begin{equation}
\sum_{\rm col}|{\cal A}(1_{\bar{b}}^{h_1},2_b^{h_2},3_{\bar
q}^{h_3},4_q^{h_4})|^2=  g_S^4 \left(\frac{y_b}{\sqrt 2}\right)^2 (N_c^2-1)
|A|^2\;.
\end{equation}

The four-$b$-quark subprocesses in Figs.~\ref{fig:real}(c) and (d) may be
obtained from the helicity amplitudes for the unphysical subprocess $0 \to
{\bar b} b {\bar b} b h$.  These helicity amplitudes may be obtained from the
four-quark amplitudes above by subtracting a term with one pair of the
identical quarks exchanged,
\begin{eqnarray}
{\cal A}(1_{\bar{b}}^{h_1},2_b^{h_2},3_{\bar b}^{h_3},4_b^{h_4})= (g_S
\sqrt{2})^2\frac{y_b}{\sqrt 2} \frac12  \left[ \left(
\delta^{\bar{\imath}_1}_{i_4}\delta^{\bar{\imath}_3}_{i_2} - \frac{1}{N_c}
\delta^{\bar{\imath}_1}_{i_2}\delta^{\bar{\imath}_3}_{i_4} \right)  A- \left(
\delta^{\bar{\imath}_1}_{i_2}\delta^{\bar{\imath}_3}_{i_4} - \frac{1}{N_c}
\delta^{\bar{\imath}_1}_{i_4}\delta^{\bar{\imath}_3}_{i_2} \right)A^{\rm ex}
\right]\;,
\end{eqnarray}
where
\begin{eqnarray}
&&A=A(1_{\bar{b}}^{h_1},2_b^{h_2},3_{\bar
b}^{h_3},4_b^{h_4})\\
&&A^{\rm ex}=\delta_{h_2 h_4}A(1_{\bar{b}}^{h_1},4_b^{h_4},3_{\bar
b}^{h_3},2_b^{h_2})+ \delta_{h_1 h_3} A(3_{\bar{b}}^{h_3},2_b^{h_2},1_{\bar
b}^{h_1},4_b^{h_4})\;.
\end{eqnarray}
The four independent helicity configurations are given by
\begin{eqnarray} &&A(1_{\bar{b}}^+,2_b^+,3_{\bar
b}^+,4_b^-)=f(1,2,3,4)+f(2,1,3,4)\\
&&A(1_{\bar{b}}^+,2_b^+,3_{\bar
b}^-,4_b^+)=f(1,2,4,3)+f(2,1,4,3)\\
&&A(1_{\bar{b}}^+,2_b^-,3_{\bar
b}^+,4_b^+)=f(3,4,1,2)+f(4,3,1,2)\\
&&A(1_{\bar{b}}^-,2_b^+,3_{\bar b}^+,4_b^+)=f(3,4,2,1)+f(4,3,2,1)\;.
\end{eqnarray}
Squaring and summing over colors gives
\begin{equation}
\sum_{\rm col}|{\cal A}(1_{\bar{b}}^{h_1},2_b^{h_2},3_{\bar
q}^{h_3},4_q^{h_4})|^2= g_S^4 \left(\frac{y_b}{\sqrt 2}\right)^2 (N_c^2-1)
\left( |A|^2+ |A^{\rm ex}|^2 + \frac{2}{N_c} {\rm Re}(A^{\rm ex} A^*) \right)
\;.
\end{equation}

\section*{Appendix~C}%
\label{sec:appc}

In this appendix we list the scalar integrals which result from the
Passarino-Veltman tensor reduction of the one-loop diagrams of
Fig.~\ref{fig:virtual}, as given in Eq.~(\ref{virtual}). We define
$(d=4-2\epsilon)$
\begin{eqnarray}
&&B_0(p_1^2;m_0^2,m_1^2) \equiv \mu^{2\epsilon} \int \frac{{\rm
d}^dk}{(2\pi)^d} \frac{1}{[k^2
-m_0^2][(k+p_1)^2-m_1^2]}\\[10pt]
&&C_0(p_1^2,p_2^2,p_{12}^2; m_0^2,m_1^2,m_2^2) \equiv \nonumber\\
&&\;\;\mu^{2\epsilon} \int \frac{{\rm d}^dk}{(2\pi)^d} \frac{1}{[k^2
-m_0^2][(k+p_1)^2-m_1^2][(k+p_1+p_2)^2-m_2^2]}\qquad\qquad\\[10pt]
&&D_0(p_1^2,p_2^2,p_3^2,p_4^2,p_{12}^2,p_{23}^2; m_0^2,m_1^2,m_2^2,m_3^2)\equiv
\nonumber \\
&&\hspace*{-1cm}\mu^{2\epsilon} \int \frac{{\rm d}^dk}{(2\pi)^d} \frac{1}{[k^2
-m_0^2][(k+p_1)^2-m_1^2][(k+p_1+p_2)^2-m_2^2][(k+p_1+p_2+p_3)^2-m_3^2]}
\end{eqnarray}
where $p_{ij}^2=(p_i+p_j)^2$, and
\begin{equation}
c_\Gamma \equiv (4 \pi)^\epsilon \frac{ \Gamma(1+\epsilon)
\Gamma^2(1-\epsilon)}{\Gamma(1-2 \epsilon)}\;.
\end{equation}
The scalar integrals needed are ($s$ and $t$ are generic invariants here)
\begin{eqnarray}
 B_0(0;0,0)  &=& 0\\
 B_0(s;0,0)  &=&
  \frac{i c_\Gamma}{16 \pi^2} \left(\frac{1}{\epsilon}  + 2
  - \ln\frac{-s}{\mu^2}\right) \qquad s<0  \\
C_0(0,0,s;0,0,0) \equiv  \frac{i C(s)}{s} &=& \frac{i c_\Gamma}{16 \pi^2}
        \frac{1}{s} \left(\frac{1}{\epsilon^2} -
        \frac{1}{\epsilon} \ln  \frac{-s}{\mu^2} +
        \frac{1}{2} \ln^2 \frac{-s}{\mu^2}\right) \quad s<0 \\[10pt]
C_0(m_h^2,0,s;0,0,0)  &\equiv&  \frac{i C'(s)}{m_h^2-s} \nonumber \\
&=& \frac{i c_\Gamma}{16 \pi^2}
        \frac{1}{m_h^2-s} \left[
        \frac{1}{\epsilon} \ln  \frac{-s}{-m_h^2} -
        \frac{1}{2} \left( \ln^2 \frac{-s}{\mu^2} - \ln^2 \frac{-m_h^2}{\mu^2}
        \right) \right]\nonumber\\
        &&  \quad s,m_h^2<0 \\[10pt]
D_0(0,0,0,m_h^2,s,t;0,0,0,0) &\equiv& \frac{i D(s,t)}{st} \nonumber\\
&=& \frac{i c_\Gamma}{16 \pi^2} \left( \frac{\mu^2}{-m_h^2} \right)^\epsilon
 \frac{2}{s t}  \left[
        \frac{1}{\epsilon^2} -
        \frac{1}{\epsilon} \left(
        \ln \frac{-s}{-m_h^2}+
        \ln \frac{-t}{-m_h^2}\right) \right.\nonumber\\
        &&+ \left.
        \frac12
        \left( \ln^2  \frac{-s}{-m_h^2}+
        \ln^2 \frac{-t}{-m_h^2} \right)+
        {\rm R}\left(\frac{-s}{-m_h^2},\frac{-t}{-m_h^2}\right)
         \right]\\[10pt]
&& \quad m_h^2<s,t<0 \nonumber \\
&& \hspace*{-6cm}{\rm R}(x,y)= \ln x \ln y -\ln x \ln (1-x) - \ln y \ln(1-y)
+\frac{\pi^2}{6} -{\rm Li_2} (x) -{\rm Li_2} (y) \;.
\end{eqnarray}
The analytic continuation of the above results to the physical region is
accomplished through the use of
\begin{equation}
\ln (-s- i \eta ) = \ln |s| - i \pi \Theta(s) \;,
\end{equation}
where $s$ is a generic invariant (including $m_h^2$) and $\eta$ is a small
positive number and, for the dilogarithms with arguments greater than unity,
by means of
\begin{equation}
{\rm Re}\left[ {\rm Li_2} (x) \right]= - {\rm Li_2} \left(\frac{1}{x}\right )+
\frac{\pi^2}{3} - \frac12 \ln^2 x \;.
\end{equation}

\section*{Appendix~D}%
\label{sec:appd}

In this appendix we derive the QCD counterterm for the Yukawa coupling in the
$\overline{\rm MS}$ renormalization scheme \cite{Braaten:1980yq}.  We consider
two different regularization schemes (RS): conventional dimensional
regularization (CDR) and modern dimensional reduction (DR).

The Yukawa coupling and the quark mass arise from a common term in the
Lagrangian,
\begin{equation}
{\cal L}= - \frac{y_0}{\sqrt 2} \overline Q Q (h+v)\;, \label{Lmass}
\end{equation}
where $y_0$ is the bare Yukawa coupling, $h$ is the physical Higgs-boson field,
and $v=(\sqrt 2 G_F)^{-1/2}$ $\approx 246$ GeV is the vacuum-expectation value
of the Higgs-doublet field.  It is evident from Eq.~(\ref{Lmass}) that the bare
quark mass is related to the bare Yukawa coupling by $m_0 = y_0v/\sqrt 2$.

The quark mass and Yukawa coupling receive corrections at one loop in QCD. We
express the bare parameters in terms of the $\overline{\rm MS}$ values and a
counterterm,
\begin{eqnarray}
&&y_0=\mu^\epsilon y(1+\delta y_{\rm RS})\\
&&m_0=\overline{m}+\delta m_{\rm RS}\;, \label{cterm}
\end{eqnarray}
where the subscript on the counterterm indicates that it depends on the
regularization scheme (RS).  The parameter $\mu$ is the 't Hooft mass,
introduced to keep the renormalized Yukawa coupling dimensionless in
$d=4-2\epsilon$ dimensions.  The Higgs vacuum-expectation value does not
receive a correction at one loop in QCD.  From the above equations and the
relation $m_0 = y_0v/\sqrt 2$ we find that the mass and Yukawa-coupling
counterterms are related by $\delta y_{\rm RS}=\delta m_{\rm
RS}/\overline{m}$. Thus we may obtain the Yukawa-coupling counterterm from the
mass counterterm.

The one-loop quark propagator is given by
\begin{equation}
\frac{i}{\hat{p}-m_0+\Sigma_{\rm RS}(\hat{p})} =
\frac{i}{\hat{p}-\overline{m}-\delta m_{\rm RS}+\Sigma_{\rm RS}(\hat{p})}\;,
\end{equation}
where $i\Sigma_{\rm RS}(\hat{p})$ is the one-loop quark self energy.  Since it
is ultraviolet divergent, it depends on the regularization scheme. The
position of the pole in the propagator at one loop is\footnote{At one loop, the
mass $m$ in the argument of the quark self energy may be regarded as the pole
mass or the $\overline{\rm MS}$ mass.}
\begin{equation}
m_{\rm pole}= \overline{m}+\delta m_{\rm RS} - \Sigma_{\rm RS}(m)\;.
\label{mpole}
\end{equation}
The $\overline{\rm MS}$ mass is defined via
\begin{equation}
\delta m_{\rm CDR}= \Sigma(m)|_{\rm div}\;, \label{dmmsbar}
\end{equation}
where $\Sigma(m)|_{\rm div}$ is the divergent part, proportional to
$c_\Gamma/\epsilon$, of the quark self energy (which is the same in CDR and
DR). Since the pole mass is a physical quantity,\footnote{The quark pole mass
is a physical quantity within perturbation theory, which suffices for our
purposes. However, it is unphysical once nonperturbative QCD is taken into
account \cite{Beneke:1994sw,Bigi:1994em}.} independent of the regularization
scheme, Eqs.~(\ref{mpole}) and (\ref{dmmsbar}) yield
\begin{equation}
\delta m_{\rm DR} = \Sigma(m)|_{\rm div} + \Sigma_{\rm DR}(m) - \Sigma_{\rm
CDR}(m)\;. \label{dmdr}
\end{equation}

The one-loop quark self energy (in 't Hooft-Feynman gauge) is
\begin{eqnarray}
 i \Sigma_{\rm RS}(\hat p) &=&   \int\frac{d^dk}{(2\pi)^d} \left(ig_S\mu^\epsilon
\gamma^\nu T^a \right) \frac{i}{\hat{p}+\hat{k}-m} \left(ig_S\mu^\epsilon
\gamma^\mu T^a \right)
\frac{-ig_{\mu\nu}}{k^2} \nonumber\\
&=& - g_S^2 C_F \left\{
\begin{array}{cl}
\left[(-2+2\epsilon ) \hat{p} + (4-2\epsilon )m \right] B(p^2)
+ (-2+2\epsilon )\hat{p} A(p^2) & {\rm in \; CDR}\\[10pt]
\left( -2 \hat{p} +  4m \right) B(p^2) -2 \hat{p} A(p^2) & {\rm in \; DR}
\end{array}
\right. \label{eq:self}
\end{eqnarray}
where $C_F=(N_c^2-1)/2N_c=4/3$ and
\begin{eqnarray}A(p^2) p^\alpha &\equiv& \mu^{2\epsilon}\int\frac{d^dk}{(2\pi)^d}
\frac{k^\alpha}{k^2 \left[(p+k)^2-m^2 \right]}\\
B(p^2) &=& B_0(p^2;0,m^2)\;.
\end{eqnarray}
Using
\begin{eqnarray}
A(m^2) &=& -\frac{i}{16 \pi^2} \left(\frac{\mu^2}{m^2}\right)^\epsilon c_\Gamma
 \left( \frac{1}{2 \epsilon} +\frac{1}{2}\right) \\
B(m^2) &=& \frac{i}{16 \pi^2} \left(\frac{\mu^2}{m^2}\right)^\epsilon c_\Gamma
\left( \frac{1}{\epsilon} +2 \right)\;,
\end{eqnarray}
gives
\begin{equation}
\Sigma_{\rm RS}(m)=-\frac{\alpha_S}{4 \pi} C_F
\left(\frac{\mu^2}{m^2}\right)^\epsilon  c_\Gamma  \left[ \frac{3}{\epsilon}+4+
\delta_{\rm RS} \right]m \;,\label{sigma}
\end{equation}
where $\delta_{\rm CDR}=0$ and $\delta_{\rm DR}=1$. The counterterm is then
obtained from Eqs.~(\ref{dmmsbar}) and (\ref{dmdr}),
\begin{equation}
\delta m_{\rm RS} = -\frac{\alpha_S}{4 \pi}  C_F c_\Gamma \left[
\frac{3}{\epsilon}{+\delta_{\rm RS}} \right]m\;. \label{deltamrs}
\end{equation}
Using $\delta y_{\rm RS}=\delta m_{\rm RS}/\overline{m}$ then gives
\begin{equation}
\delta y_{\rm RS} = -\frac{\alpha_S}{4 \pi}  C_F c_\Gamma \left[
\frac{3}{\epsilon}{+\delta_{\rm RS}} \right]\;. \label{deltayrs}
\end{equation}
This result is independent of the gauge chosen.

The $\overline{\rm MS}$ counterterm for the strong coupling,
$g_S^0=\mu^\epsilon g_S(1+\delta g_S^{\rm RS})$, analogous to
Eq.~(\ref{deltayrs}), is \cite{Kunszt:1994sd}
\begin{equation}
\delta g_S^{\rm RS} = \frac{\alpha_S}{4 \pi} c_\Gamma
\left[-\frac{b_0}{\epsilon} + \frac{C_A}{6} \delta_{\rm RS} \right]\;,
\label{asren}
\end{equation}
where $b_0=\beta_0/2=(11/6)C_A - (2/3) T_F n_f$, $C_A=N_c=3$, $T_F=1/2$, $n_f$
is the number of light quarks, and $\delta_{\rm CDR}=0$, $\delta_{\rm DR}=1$.

The relation between the pole mass and $\overline{\rm MS}$ mass may also be
obtained from Eqs.~(\ref{mpole}), (\ref{dmmsbar}), and (\ref{sigma}),
\begin{equation}
m_{\rm pole}= \overline{m}(\mu) \left[1+ \frac{\alpha_S}{4 \pi} C_F
\left(4+3\ln\frac{\mu^2}{m^2}\right) \right]\;. \label{MSpole}
\end{equation}

\section*{Appendix~E}%
\label{sec:appe}

In this appendix we discuss two of the checks performed on our calculation of
the virtual correction presented in Section~\ref{sec:virtual} and Appendix~C.
We checked that the structure of the infrared (soft and collinear) divergences
is as expected from the dipole-subtraction method \cite{Catani:1996vz}.  We
also performed the calculation in both conventional dimensional regularization
(CDR) and modern dimensional reduction (DR) and verified the scheme
independence of our results.

The structure of the divergences for $gb \to hb$ at one loop is
\begin{eqnarray}
&& \hspace{-1cm} 2 {\rm Re}\left( {\cal A}_1 {\cal A}_0^* \right) |_{\rm div}
= |{\cal A}_0|^2
      \frac{\alpha_S}{2 \pi} c_\Gamma \left( \frac{\mu^2}{m_h^2} \right)^\epsilon
\nonumber \\
&&\hspace{-1cm}
           \times\left[-\frac{1}{\epsilon^2}  (C_A+ 2 C_F) +
            \frac{1}{\epsilon}  \left( C_A \left(
             \ln \frac{s}{m_h^2}
             +\ln \frac{-t}{m_h^2} \right)
             - (C_A -2 C_F) \ln \frac{-u}{m_h^2} \right)  \right]\;,
\label{eq:beforeUVsub}
\end{eqnarray}
where $C_F=(N_c^2-1)/2N_c=4/3, C_A=N_c=3$ and where ${\cal A}_0$ is the tree
amplitude given in Eqs.~(\ref{treeapp}) and (\ref{treehel}), and ${\cal
A}_1\equiv \epsilon_\mu^{h_3*}(p_3){\cal A}_1^\mu$ where ${\cal A}_1^\mu$ is
the one-loop amplitude given in Eq.~(\ref{virtual}).  This expression contains
both infrared and ultraviolet divergences.  The latter are removed by
renormalizing the Yukawa and strong couplings using the counterterms given in
Eqs.~(\ref{deltayrs}) and (\ref{asren}), respectively.  This leaves the
infrared-divergent expression
\begin{eqnarray}
&&\left.\left[2 {\rm Re}\left( {\cal A}_1 {\cal A}_0^* \right) + |{\cal
A}_0|^22(\delta y_{\rm RS} + \delta g_S^{\rm RS})\right] \right|_{\rm div} =
|{\cal A}_0|^2
      \frac{\alpha_S}{2 \pi} c_\Gamma  \left( \frac{\mu^2}{m_h^2} \right)^\epsilon
           \left[-\frac{1}{\epsilon^2}  (C_A+ 2 C_F)
           \nonumber \right.\\ &&\hspace*{.5cm} +\left.
            \frac{1}{\epsilon}  \left( C_A \left(
             \ln \frac{s}{m_h^2}
             +\ln \frac{-t}{m_h^2} \right)
             - (C_A -2 C_F) \ln \frac{-u}{m_h^2}
             - b_0  -3 C_F \right)  \right]\;,
\end{eqnarray}
which has the structure expected from the dipole-subtraction method
\cite{Catani:1996vz}.

We also calculate the relation between the virtual amplitude in CDR and DR. We
find
\begin{eqnarray}
    2 {\rm Re}\left( {\cal A}_1 {\cal A}^*_0 \right) |_{\rm CDR}=
    2 {\rm Re}\left( {\cal A}_1 {\cal A}^*_0 \right) |_{\rm DR}
    -  |{\cal A}_0|^2  \frac{\alpha_S}{2 \pi}  2 C_F \;.
\label{cdr}
\end{eqnarray}
The above relation is consistent with the set of rules given in
Ref.~\cite{Kunszt:1994sd}, augmented by the regularization-scheme-dependent
renormalization of the Yukawa coupling given in Eq.~(\ref{deltayrs}).

%

\end{document}